\documentclass{an}
\usepackage{graphicx}
\usepackage{times}
\usepackage{txfonts}

\Volume{00}              
\Year{2010}              
\Month{00}               
\Pagespan{000}{111}      

\def\gee{ \, \lower 1mm\hbox{$\,{\buildrel > \over{\scriptstyle\scriptstyle\sim} }\displaystyle \,$}}
\def\lee{ \, \lower 1mm\hbox{$\,{\buildrel < \over{\scriptstyle\scriptstyle\sim} }\displaystyle \,$}}
\def\|{\partial}
\def\o {\over}
\def\Oo {\displaystyle}
\def\varkappa {{\scriptstyle\partial}\! e}

\begin{document}

\lhead[A. Khoperskov et al.: Numerical modelling of disc galaxies]{\thepage}
\rhead[Astron. Nachr./AN~{\bf XXX} (200X) X]{\thepage}
\headnote{Astron. Nachr./AN {\bf 32X} (200X) X, XXX--XXX}

\title{Numerical modelling
of the vertical structure and dark halo parameters in disc
galaxies }

\author{A. Khoperskov\inst{1}, D. Bizyaev\inst{2,3}, N. Tiurina\inst{2},
M. Butenko\inst{1}}

\institute{(1) Volgograd State University, Volgograd, 400062, Russia \\
(2) Sternberg Astronomical Institute, Moscow, 119992, Russia \\
(3) Apache Point Observatory and NMSU, Sunspot, NM,USA
}

\date{Accepted ???? ???? ??. Received ???? ???? ??;
in original form ???? ???? ??}

\begin{abstract}
\noindent \,\,
The non-linear dynamics of bending instability and vertical
structure of a galactic stellar disc embedded into a spherical
halo are studied with N-body numerical modelling.
Development of the bending instability in stellar galactic disc is
considered as the main factor that increases the disc thickness.
Correlation between the disc vertical scale height and the halo-to-disc mass
ratio is predicted from the simulations.
The method of assessment of the
spherical-to-disc mass ratio for edge-on spiral galaxies with a
small bulge is considered. Modelling of eight
edge-on galaxies: NGC~891, NGC~4738, NGC~5170, UGC~6080,
UGC~7321, UGC~8286, UGC~9422 and UGC~9556 is performed.
Parameters of stellar discs, dark haloes and bulges are estimated. The
lower limit of the dark-to-luminous mass ratio in our galaxies is of
the order of 1 within the limits of their stellar discs.
The dark haloes dominate by mass in the
galaxies with very thin stellar discs (NGC~5170, UGC~7321 and
UGC~8286).
\end{abstract}


\correspondence{dmbiz@apo.nmsu.edu}

\maketitle

\section{Introduction}

The luminous matter reveals only a part of the
total mass in galaxies. Additional fraction of non-luminous matter
is required to explain rotation curves (RC
hereafter) of spiral galaxies.
Fitting of galactic
rotation curves does not strictly require a lot of dark
matter. Obviously, a modest dark halo with 1/4 of the disc's mass
is enough to explain a flat RC at the radii range
$r\simeq (2-4)\cdot L$ (\cite{Albada}), where $L$ is the disc scale
length.
Numerical modelling of disc galaxies without a massive spheroidal
subsystem shows that their stellar discs heat up
significantly during the evolution, and finally achieve the
equilibrium state with too high velocity dispersion
(\cite{Ostriker, Carlberg-Sellwood1985,
Athanasoula-Sellwood-1986, Bottema1997, Fuchs-Linden-1998, AVH2003}).
To explain the low ratio of the stellar velocity dispersion to the circular
velocity $c^{obs}/V_c$, which is observed in many galaxies at $r>2L$,
a spherical subsystem as massive as the the stellar disc or even
more has to be incorporated into the models.
More additional arguments for the massive spherical haloes come
from observations of galactic polar rings and from long flat RCs
observed far outside the limits of stellar discs
(\cite{Whitmore, Reshetnikov2000, Iodice, Combes}).

Mass of the spherical subsystem in a galaxy
can be determined from its rotation with quite high level of
uncertainty. Incorporation of radial distribution
of the stellar velocity dispersion into the modelling helps to
decrease the ambiguities and allows to constrain better the
range of parameters of galactic
subsystems (\cite{Bottema1993, Bottema1997, Fux-1997, AVH2001}).
Unfortunately, observations of the stellar velocity dispersion in
galactic discs require a sensitive spectroscopy and long
integration time, so the dispersion still can be estimated for a limited
number of galaxies. An alternative method of including the
velocity dispersion into the numerical modelling is to utilize
the disc thickness that, in a contrary, is
easy to observe in many edge-on galaxies. Relation
between the vertical scale height, local disc surface density and
spheroidal subsystem mass $M_s$ enables us to include the
disc thickness into the modelling. At the same time, if the
velocity dispersion is close to that required for the
marginal disc stability, the thickness of the stellar disc is tightly
connected with the spherical-to-disc components mass ratio
(\cite{Zasov1991, AVH20012, Sotnikova06}).

One of important problems in the physics of galaxies is the
identification of the main factor that increases the random velocity
from $6\div 10$~km/s (young population) to the values of $\sim
20\div 150$~km/s (observed in old stellar population). Besides
such factors of heating as the scattering of stars by giant
molecular clouds (\cite{Spitzer-Schwarzschild-1951,
Spitzer-Schwarzschild-1953}) and in the spiral arms
(\cite{Jenkins}), the bending instability is the most important
mechanism of the vertical velocity dispersion $c_z$ growth in
disc galaxies (\cite{Polyachenko77, Zasov1991, AVH20012,
Sotnikova, Sotnikova05}).

The bending instability in galactic discs has been
considered with the help of N-body simulations. Essential
results were obtained when the tidal interaction
was taken into account (\cite{Hernq,Weinberg,Velazquez, Mayer, Bailin}).
Tidal forces are generated by the disc bending and increase the disc
thickness. This is in agreement with observations of the
probability of bending and warps in isolated galaxies and group
objects (\cite{Reshetnikov2002}).
The emergence of the bending instabilities was
investigated by \cite{Raha, Sellwood, Sellwood-Merritt-1994,
Patsis-2002, Griv-Yuan-Gedalin-2002, Binney}.
A model of the galaxy NGC~2146 that involved the bending instability
was considered by \cite{Griv}.
The bending instability reveals itself on much longer time scale,
typically 10$^9$ years (\cite{Sotnikova}) in the comparison with
the gravitational instability in the disc plane.
\cite{Hunter} explained the global warps
observed in discs of some isolated galaxies by bending instabilities.

The disc scale height in galaxies depends on the vertical velocity
dispersion $c_z(r)$. The latter is dependent of the radial
velocity dispersion $c_r(r)$. A discussion about the
relation between $c_r$ and $c_z$ was opened firstly by
\cite{Toomre1966}.
Considering a simplified model of infinitely thin
uniform self-gravitating layer, (\cite{Polyachenko77}) found
an essential condition for a stellar system stability
against the small-scale bending perturbations: $c_z/c_r
\ge 0.37$. The dynamics of the bending instabilities was
also considered by \cite{Araki-1986}, where a non-uniform
volume density distribution in the $z$-direction was taken
into account. The conclusion about a lower value of the
critical ratio $c_z/c_r \ga 0.3$ was made in that
paper (see also discussion in \cite{Merritt}).

In this paper we point our special attention to the radial distribution
of the disc thickness and ratio $c_z/c_r$ that are necessary to provide
the disc stability against different kinds of bending
perturbations. We mostly consider isolated late type galaxies without a
prominent bulge.
The presence of bulge
would complicate the analysis of bending instability.
If there is no bulge in a galaxy, one can restore the internal part of
rotation curve assuming that it is defined by the disc
component only in the inner ($r\la 2L$) regions (\cite{Zasov03}).
However, we also develop our modelling for two well-studied disc galaxies
with significant bulge component because we can incorporate observations of
of the stellar velocity dispersion for these galaxies.

In \S~2 we describe our N-body model. Evolution and features of the bending
perturbations are considered in \S~3. In \S~4 we model
structural parameters and rotation curves of eight
edge-on galaxies and evaluate their best fitting parameters focusing
at dark halo-to-disc mass ratio.

\section{Dynamical modelling of stellar discs in galaxies}

Our N-body modelling is based on the numerical integration of the
motion equations for $N$ gravitationally interacting particles
as it was described in detail by \cite{AVH2003}.
This system of collisionless particles has a shape of a disc
(and, optionally, bulge) embedded into a spherical dark halo.
The halo and bulge are represented by fixed potentials.
The steady
state distribution of mass in the bulge $\varrho^{(b)}$ and
halo $\varrho^{(h)}$ are defined as:

\begin{equation}
\label{densityhalo}
\varrho^{(h,b)}(r) = {\varrho_{0}^{(h,b)}\o (1+\xi^2/a_{(h,b)}^2)^k} \,,
\end{equation}

\noindent where $\xi=\sqrt{r^2+z^2}$, and $(r,z)$ are
the radial and vertical coordinates, $k$ equals to 3/2
for the bulge and to 1 for the halo. The dimensional
spatial scales for the bulge and halo are denoted as
$a_b$ and $a_h$, respectively.
Our model bulge is defined within a sphere with radius
$\xi \le r_b^{\max}$.

The initial vertical equilibrium in the disc is defined by
the Poisson equation

\begin{equation}
\label{Puasson}
{\|\o r\| r}\, \left( r{\| \Phi\o \| r} \right) + {\|^2 \Phi\o \|
z^2} = 4\pi G \, \left(\varrho + \varrho^{h} + \varrho^{b} \right) \,
\end{equation}

\noindent and by a balance of forces in the vertical
direction (in the first-order approximation)

\begin{equation}
\label{z-equilibrim}
c_z^2\, {\| \varrho\o \| z} = - \varrho \, {\| \Phi\o \| z} \,.
\end{equation}

\noindent Here $\Phi$ is the gravitational potential,
$\varrho$ is the disc volume density and $c_z$ is the
vertical component of velocity dispersion. The  radial
component of Jeans equation defines the rotational
velocity in the stellar disc (\cite{Valluri-1994}):

\begin{equation}
\label{VelocityRotation}
\begin{array}{l}
V^2 ~=~ \displaystyle
 V_c^2 + c_r^2\, \left\{ 1
- {c_\varphi^2\o c_r^2} + {r\o \varrho c_r^2}{ \| (\varrho c_r^2)
\o \| r} + {r\o c_r^2}{\| \langle{uw}\rangle \o \| z} \right\} \,,
\end{array}
\end{equation}


\noindent where the brackets $\langle \ldots \rangle$ denote the
mean value. The last term in the figured brackets in
(\ref{VelocityRotation}) is the chaotic part of the radial
$u$ and vertical $w$ velocity components.

The system of equations (\ref{Puasson})--(\ref{z-equilibrim}) can
be reduced to the equation for dimensionless disc density
$f(z;r)=\varrho(z;r)/\varrho(z=0;r)$, see \cite{Bahcall}:

\begin{equation}\label{Bachall}
\begin{array}{l} \displaystyle
{d^2 f \o dz^2} + 2{d\ln c_z\o dz}\, {df\o dz}-{1\o f}\, \left(
{df\o dz} \right)^2 +
 {4\pi G\o c_z^2}{\sigma\o 2 z_0}\,\times \\ \displaystyle
 \times f\, \left( f + {2\varrho_h^{eff}\o
\sigma}z_0\right)=0 \,.
\end{array}
\end{equation}

\noindent where the surface density is
$\sigma=2\varrho(z=0)\cdot z_0$, $\Oo
\varrho_h^{eff}=\varrho_h-{1\o 4\pi G r}{\| V_c^2\o \|
r}$ and the disc scale height is
$z_0=\int^{\infty}_{0}fdz$.
For the case of
$c_z={\rm const}$ and $\varrho_{h}^{eff}=0$ the
volume density profile of the disc in its vertical direction
is:

\begin{equation}
\label{cosh2}
\varrho(z)=\varrho(z=0)\cdot {\rm sech}^{2}(z/z_0) \,,
\end{equation}

\noindent where the vertical disc scale for this case
is $z_0=c_z^2/\pi G\sigma$.

At the flat part of RC, the last term of equation (\ref{Bachall})
decreases the vertical scale $z_0$ because of additional gravitational
potential. Nevertheless, this factor is not dominant and the most significant
reason for $z_0$ decreasing is that the potential of the halo
changes the value of $c_z/c_r$ required for the marginal
stability of bending perturbations.

We assume that the disc volume density can be written

\begin{equation}
\label{varrho(r,z)}
\varrho(r,z)=\varrho(0,0)\,f_r(r)\,\,f_z(z)\,,
\end{equation}

\noindent where the function $f_z(z)$ describes the vertical
density profile. We consider two functional forms for the vertical
density distribution $f_z(z)$ in our modelling: $\,\exp(-z/h_z)$ and ${\rm
sech}^{2}(z/z_0)$, where the scale heights $h_z$ and $z_0$ are functions of
time and location in the disc. The radial density profile $f_r(r)$
is assumed to be exponential: $f_r(r) = \exp(-r/L)$.

The mass $M_h$ of dark halo was calculated inside the maximum
disc radius $R_{\max}$. We assume that $R_{\max} \approx
4\,L$ in our modelling (in accordance
to \cite{Kruit1981, Kruit1982, Pohlen, Holley}).
We also designate the spherical-to-disc mass ratio in the galaxies as
$\mu$.

As the initial templates in our models, we consider axisymmetric
systems in the state of equilibrium. Initial
radial distributions of the velocity dispersions in radial ($c_r$)
and azimuthal ($c_\varphi=c_r\varkappa/2\Omega$) directions keep
the templates in gravitationally stable state.
We assume that the stellar disc population
is shaped as a quasi-stationary balanced system and that
the radial and vertical random motions of particles
keep the disc in the marginal stability
against bending waves and perturbations in the disc
plane. Then we observe the evolution of the discs
during 10--40 turns at its outer edge.

We use TREE-code approximation in the modelling
(described in detail by \cite{AVH2003})
with $N=(0.5 - 8) 10^6$. Also, in \S~3.1.1
we verify some model conclusions with independent direct
particle-particle approach (with less number of N) as well
described by \cite{AVH2003}.

For the comparison between the modelling and observations, we
incorporate both structural parameters of stellar discs and
their rotation curves at the same time. We take into account that
the observed structural parameters in edge-on galaxies are
results of integration along the line of sight.
To obtain the edge-on surface density, we integrate the volume
density of the model disc along the line of sight:

\begin{equation}\label{Density-on-sight}
\sigma(x,z)=\int\limits_{-y_0}^{y_0} \varrho(\xi , z) \, dy \ \ \ \
(\xi=\sqrt{x^2+y^2}) \,.
\end{equation}

\noindent Here $x$ is the projected distance to the centre of the disc,
$y_0=\sqrt{R_{\max}^2-x^2}$.

The similar approach is applied for the integration
of RC along the line of sight (it was described by \cite{Zasov03}).

\section{ Bending instabilities in stellar discs}

\subsection{ The global bending instability}

The linear analysis of stability against
the global bending perturbations was discussed by \cite{Polyachenko79}
and generalized by \cite{Vandervoort}.
It was shown that the large scale instability
modes with the azimuthal numbers $m = 1,2$ and $3$ generate the
instability in collisionless ellipsoidal systems in dependence of
their axes ratio, velocity dispersion and mass of their
spherical subsystems. The most important result from the
papers mentioned above is that the area of stability in
the parametric space depends on the spherical subsystem
mass. Once the halo is massive enough, the systems with
low velocity dispersion can be unstable. This gives an example
of destabilizing role of the halo. On the
other hand, the higher mass of the spherical subsystem requires
the lower ratio of $c_r/V^{\max}$ that is necessary to keep the disc
gravitationally stable. As a result, the destabilizing role of the
massive halo related to the large scale bending
perturbations may not reveal itself (\cite{Zasov1991,AVH20012}).


Our model stellar discs at the initial moment ($t=0$)
are axisymmetric, in the equilibrium state along both
the radial and vertical directions and gravitationally
stable in the disc plane. The latter is made by assigning a high
value to the radial velocity dispersion $c_r \ga
(1.5\div 4)\cdot c_T$, where the Toomre stability
parameter $Q_T=c_r/c_T$ is a function of $r$ (\cite{AVH2003,AVH2009}).
In general, our initial disc templates
can be either stable or unstable against the bending
perturbations in dependence of $c_z(r)$ distribution.

The evolution of systems with a small initial value of
$c_z/c_r$ (i.e. systems that are dynamically cold in
the vertical direction) reveals a gradual growth of the
global bending instability, which heats up the disc in
the vertical direction and increases its thickness.

\subsubsection{The axisymmetric bending mode ($m=0$)}

\begin{figure*}
\includegraphics[width=8.5cm]{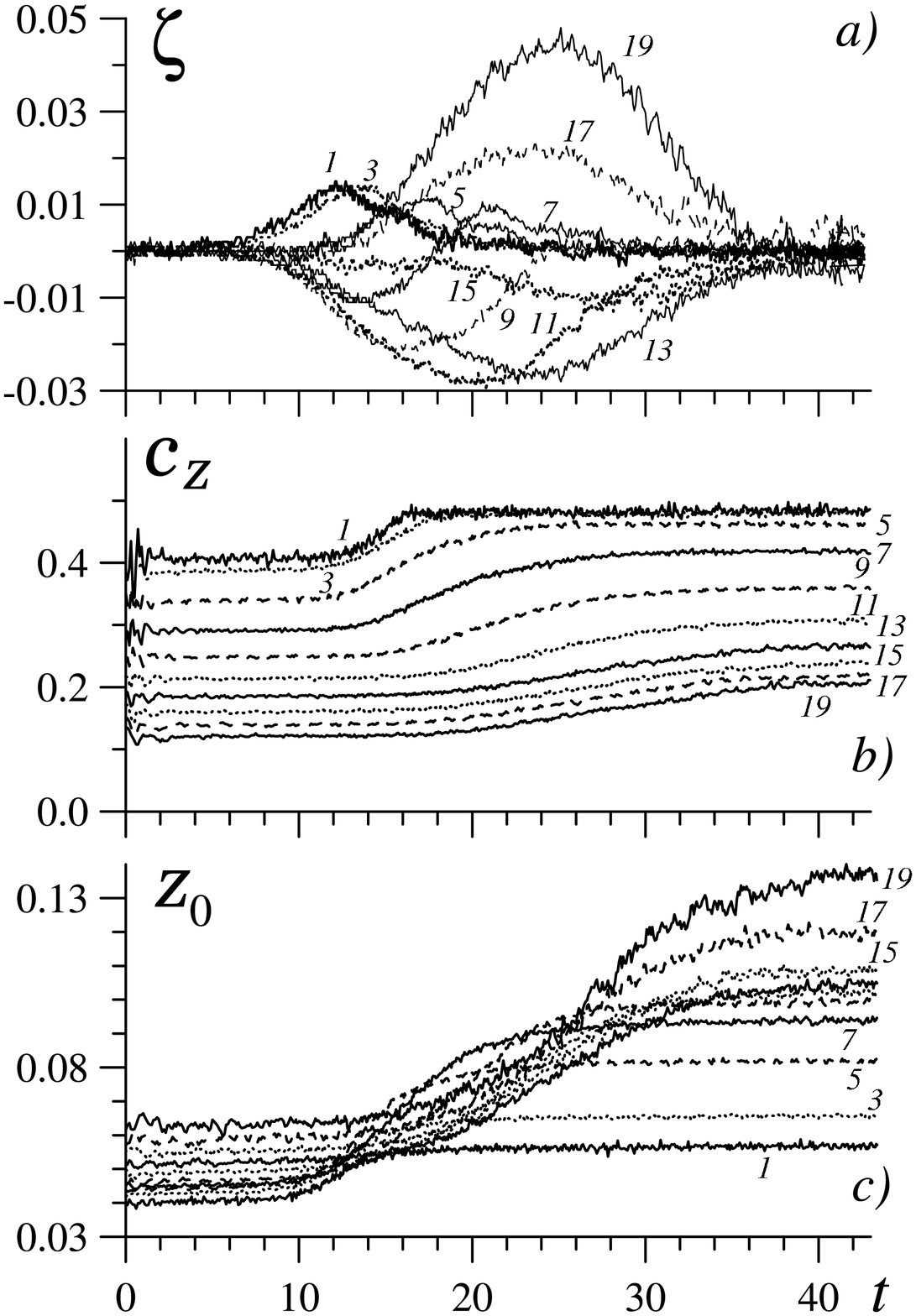}
\includegraphics[width=8.5cm]{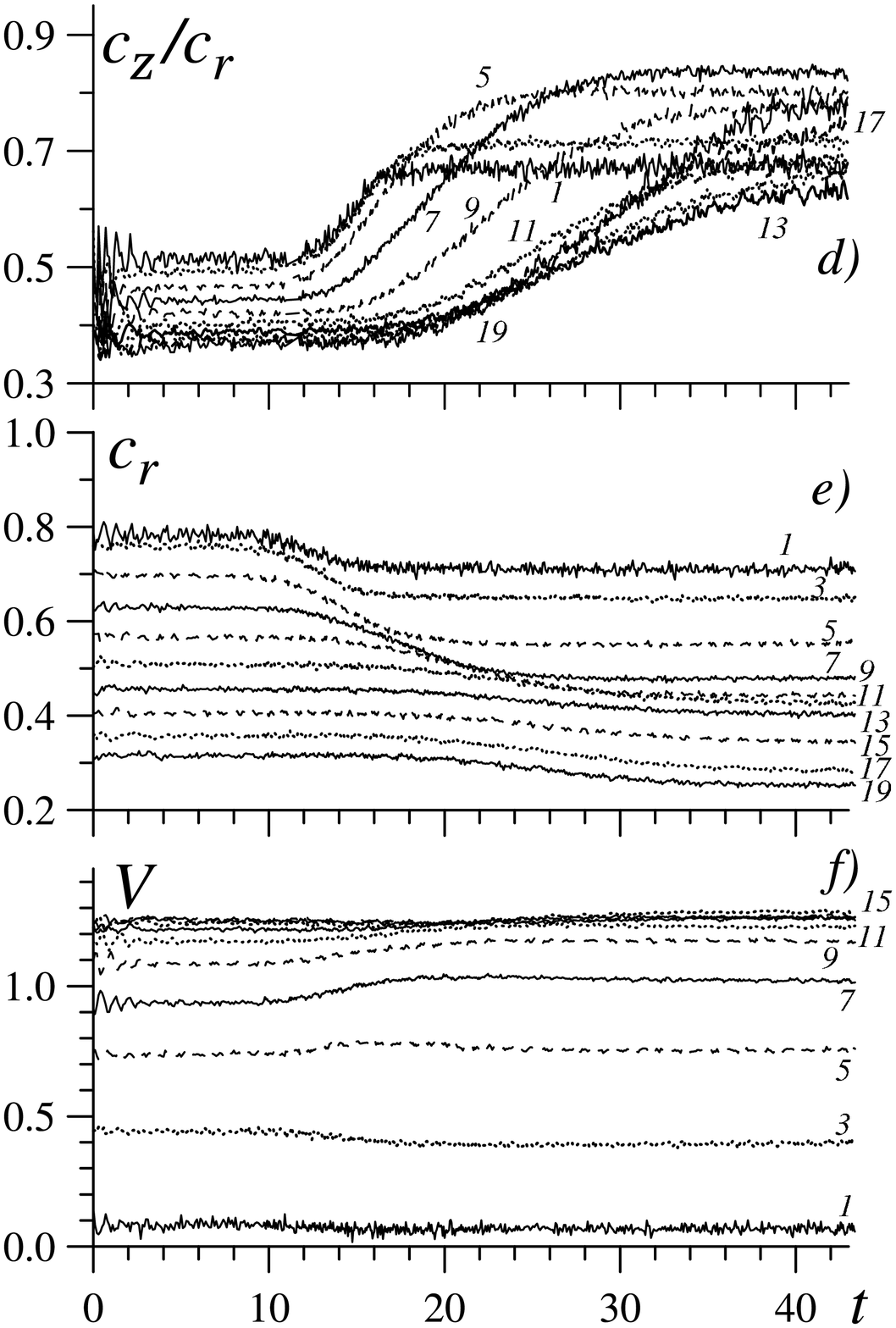}
\caption{
Evolution of the vertical structure parameters of a model stellar disc
for the case of $\mu=1$, $a=L$ when the axisymmetric
bending mode is being developed: a) the vertical coordinate of the
disc's local centre of mass $\zeta$, b) vertical component
of the velocity dispersion $c_z$, c) disc scale
height $z_0$, d) vertical-to-radial velocity
dispersions ratio $c_z / c_r$, e) $c_r$, f) rotational
velocity of particles in the disc. Different
curves are drawn for the following set of distances to the centre:
$r_j = 4L\cdot(0.05j-0.025)$, where $j$ is shown by the corresponding
curve. All the parameters are averaged along the azimuthal coordinate.
}
\label{fig1}
\end{figure*}

Development of unstable axisymmetric bending mode $m=0$ leads
to significant disc heating in the vertical direction
(the same conclusion has been made by \cite{Sellwood}).
The evolution of the vertical coordinate of the
local centre of mass $\zeta(r,t)$, vertical velocity dispersion
$c_z(r,t)$, disc scale height $z_0(r,t)$, dispersion ratio
$\alpha_z\equiv c_z(r,t)/c_r(r,t)$, radial velocity dispersion
$c_r(r,t)$ and  rotational velocity $V(r,t)$ in the stellar disc
where the axisymmetric mode is developing can be seen in Fig.~\ref{fig1}.
We fix the units hereafter assuming that $M_d = 1$, $G=1$ and
$4L=1$. In this case the circular velocity $V_c \sim 1.5$ at the
periphery of the disc and the corresponding dimensionless rotation period
is $\tau \approx 4$.

As it is seen in Fig.~\ref{fig1}, parameters of the disc stay almost
unchanged for the first 2.5 turns ($t \approx 10$ in our units).
This time the instability modes are being developed at the linear
stage of the bending instability evolution. After $t \ga 10$, the
linear stage is superseded by the non-linear development of the
bending instability, and the centre of mass $\zeta(r)$ oscillates
with larger amplitude in the $z$-direction (see Fig.~\ref{fig1}a). The
amplitude $\zeta(r)$ rapidly increases in the central regions of
the disc up to its maximum value and then decreases down to the
initial state (see curves 1--5 in Fig.~\ref{fig1}a), whereas the increasing
of $\zeta(r)$ occurs much slower at the outer parts of the disc
(curves 11--19). The rapid growth of the velocity dispersion $c_z$
and scale height $z_0$ begins after a time delay from $\zeta(r)$,
see Fig.~\ref{fig1}b and c. A flare of stellar disc emerges in
its central region and
then propagates outside, towards the periphery. The value of $c_z/c_r$
rises mostly because of $c_z$ growth, and in less power due to $c_r$
decreasing (Fig.~\ref{fig1}d and e). The azimuthal component of
the velocity dispersion $c_\varphi$ follows $c_r$, so the relation
$c_\varphi\simeq c_r \varkappa/2\Omega$ is valid almost
everywhere. This equation can be failed either in a very
thick disc ($z_0/L\ga 0.4$), or in the very central region, where
$c_r/c_\varphi < 2\Omega/\varkappa$, that means the
anisotropy is lower in those regions because of the system spherization.
Decreasing of $c_r$ and $c_\varphi$ is the result of conversion of the
kinetic energy of random motions in the disc plane to the kinetic
energy of random vertical motions. Note that development
of this bending mode takes place in the disc which is initially
axisymmetric over all parameters.

\begin{figure*}
\includegraphics[width=8.5cm]{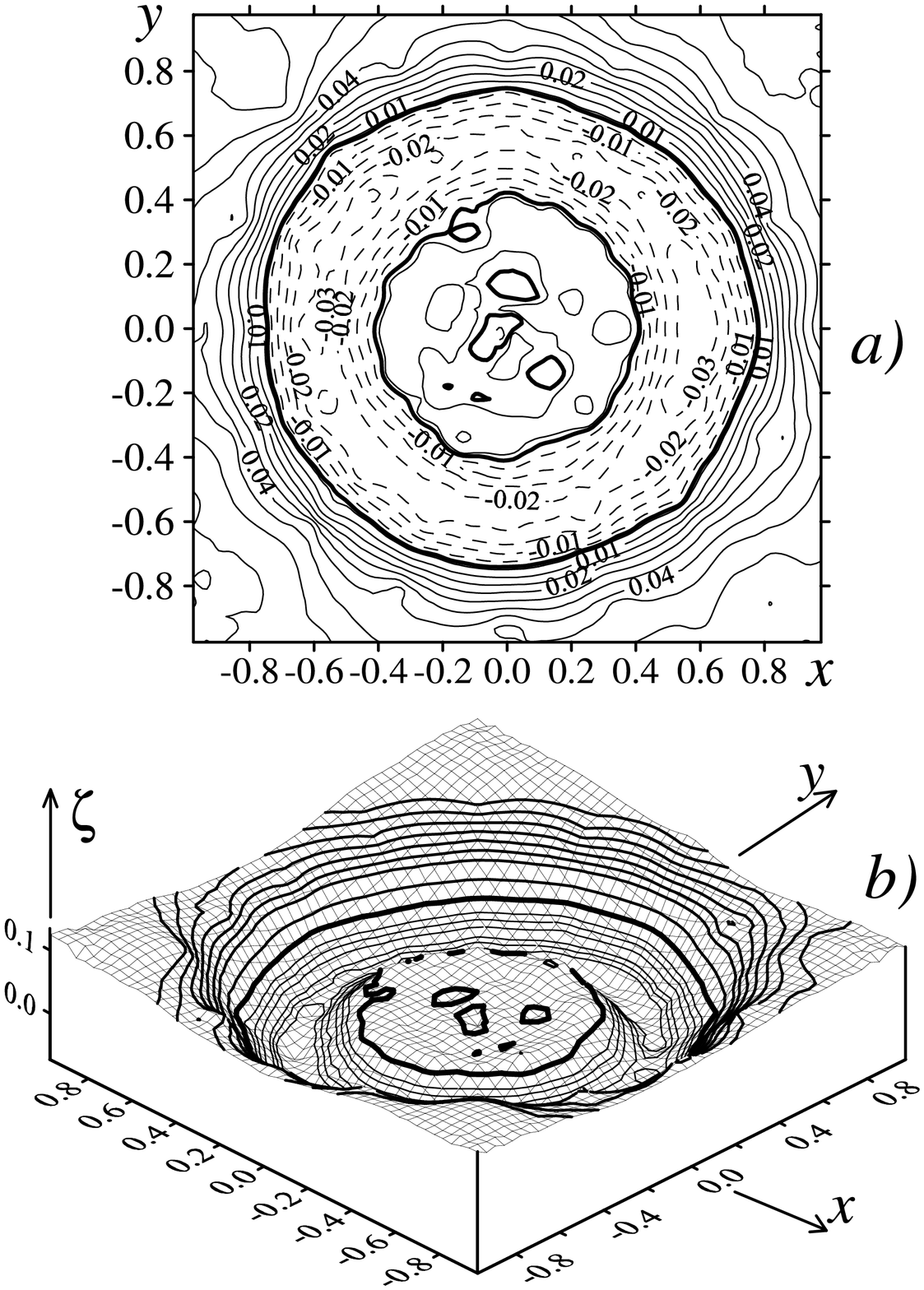}
\includegraphics[width=8.5cm]{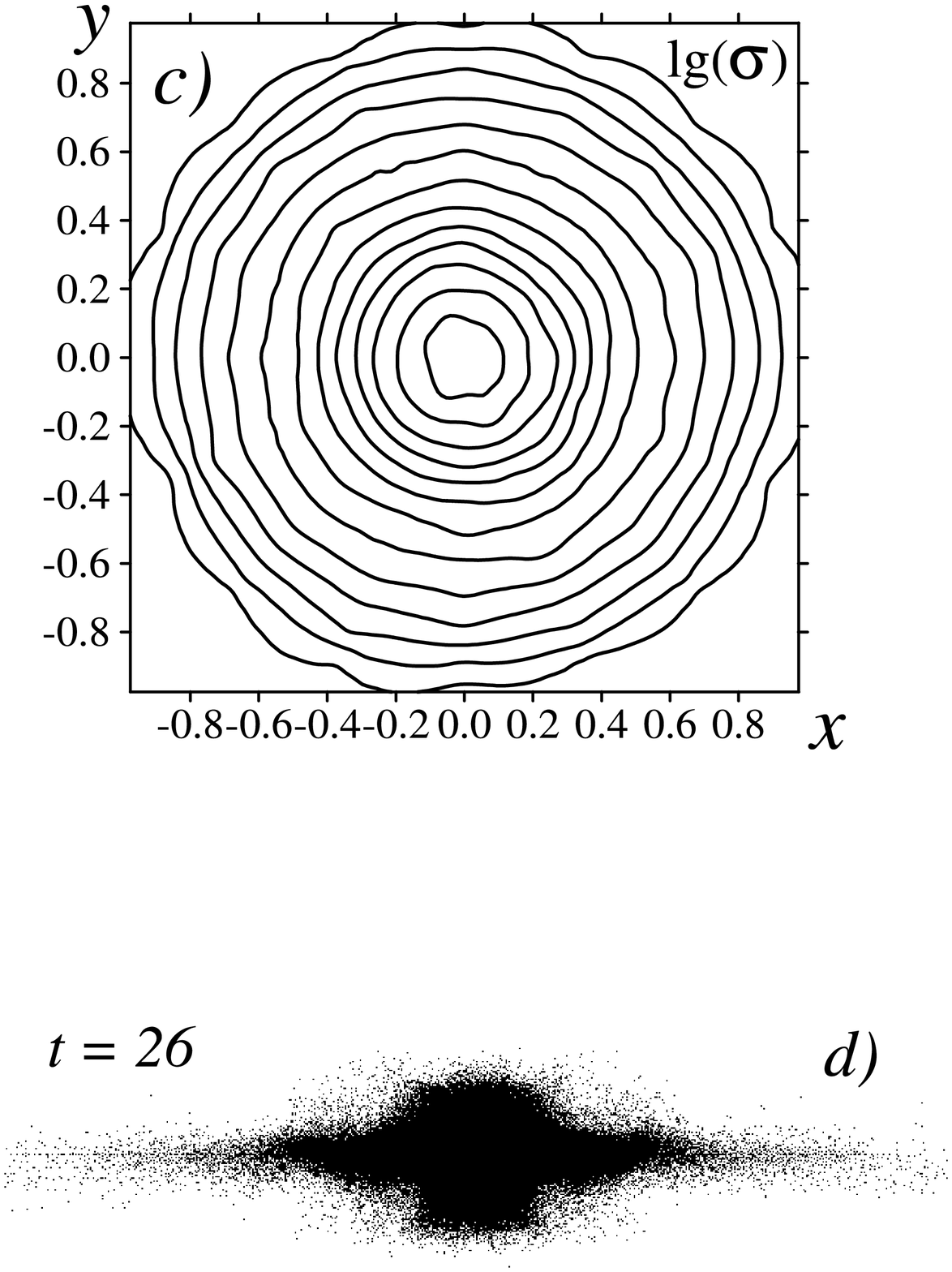}
\caption{
The shape of a model disc that experiences a development of the
bending instability at the moment $t=25.4$ for the model
shown in Fig.~\ref{fig1}. The isolines denote the
distribution $\zeta(r,\varphi)$ in the plane of the disc (a) and in
the x-y plane $\zeta(x,y)$ (b).
The bold solid line represents
$\zeta=0$. c) Isolines of logarithm of the surface density
are shown at the same moments as in the panel (a).
The axial symmetry in the surface density distribution can
be seen in the figure.
d) Edge-on view at the disc on the stage of the global
axisymmetric mode development. The vertical/radial aspect
ratio in Figure is increased by the factor of 4.
}
\label{fig2}
\end{figure*}

The value of $\zeta$ has opposite signs at the disc centre and
periphery in the considering model (see Fig.~\ref{fig1}a). This
generates a ``mexican hat-like'' structure as a result of this
instability (Fig.~\ref{fig2}). The isolines of $\zeta$ have concentric
shape. The intermediate regions and periphery at the moment $t=25.4\simeq
6\tau$ are shifted off the disc plane toward the opposite
directions, whereas the central region ($r\la 0.4$) has returned
to its initial position, see Fig.~\ref{fig2}b. The disc vertical structure
driven by the bending mode $m=0$ is shown in Fig.~\ref{fig2}d. We can see
a formation of
box-shaped features in the disc, which is typical for all moments
of evolution when
the axisymmetric bending mode dominates. Note that this feature is
not relevant to a bar. The considered instability may be responsible
for the formation of central thickening of discs (i.e. bulges), and
for the structure of S0 galaxies (see Fig.~\ref{fig2}d)).

A gradual change of the velocity dispersions $c_z$ and
$c_r$ and rotational velocity $V$ begins when the
amplitude of the vertical oscillations $\zeta$ grows
significantly. The vertical disc heating is accompanied by
the increasing of $c_z/c_r$, but
favourable conditions for developing
the bending instability disappear at certain values
of $c_z/c_r$. As a result, new
stable and thick disc forms. The specific time for this
process depends significantly on the model parameters and
is of the order of ten turns of the disc's outer regions. If the
initial distribution of $\alpha_z = c_z/c_r$ was
subcritical, the linear stage takes longer time due to
low increment of instability, and because of the heating at
the non-linear stage was feeble. The
amplitude of the bending mode, and even the ability of its
emergence, depends significantly on the initial radial
distribution of $c_z/c_r$.

An important consequence is that the final distribution of $\alpha_z(r)$,
as a result of bending instability, depends on its initial
value $\alpha_z (r, t=0)$. After development of the global bending
instability, those thin discs that had a low value of the scale height ratio
gain larger ratio $c_z/c_r$ than it would required to keep
the discs stable. The explanation can be found in the
essentially non-linear nature of the disc heating.
As a result, the parameter $c_z/c_r$ ``jumps over'' the
``border of stability'' and comes to relaxation well above
its marginal threshold.

\begin{figure*}
\includegraphics[height=5cm]{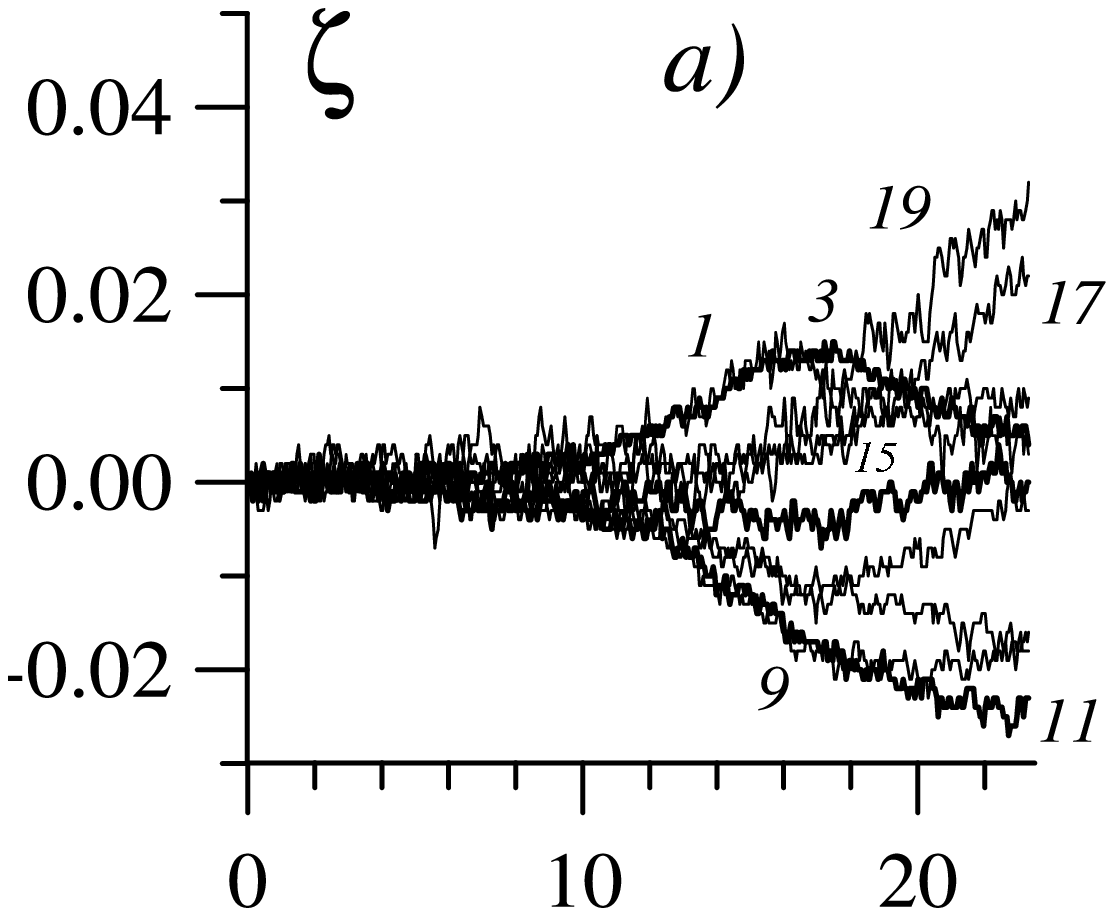}
\includegraphics[height=5cm]{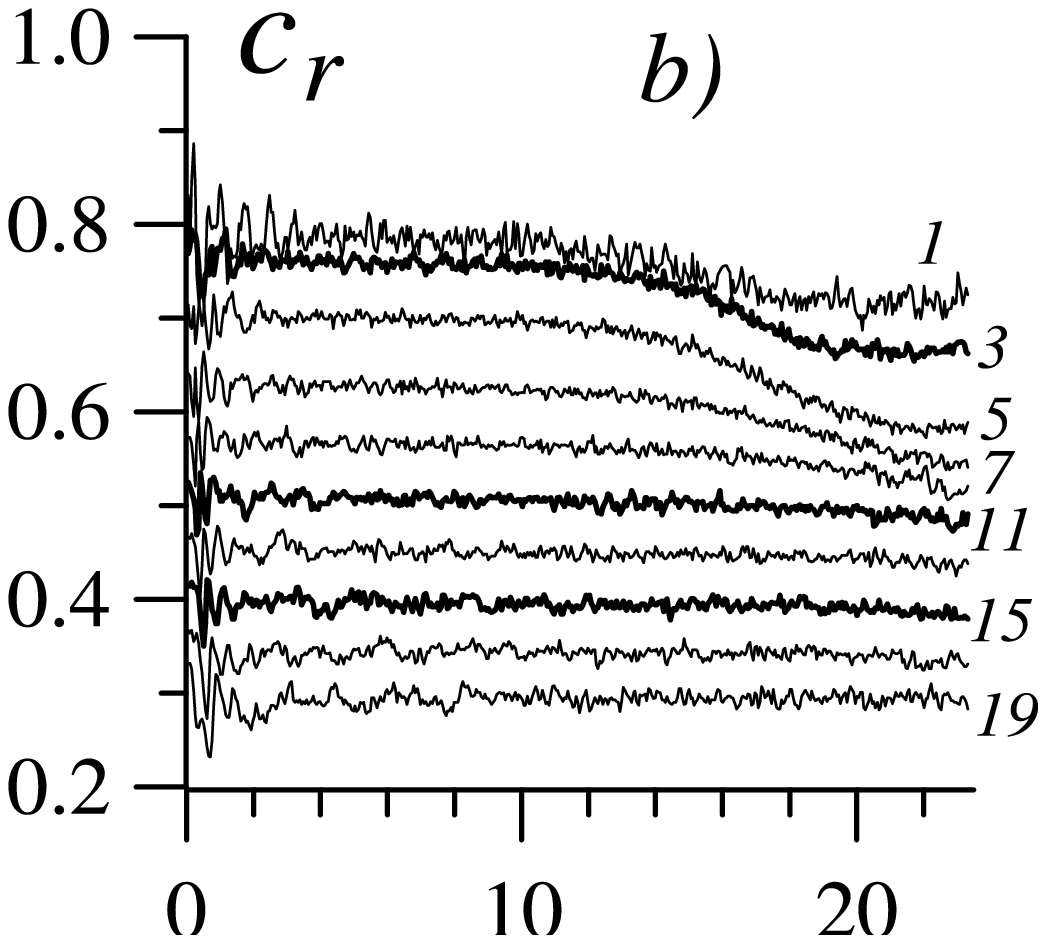}
\includegraphics[height=5cm]{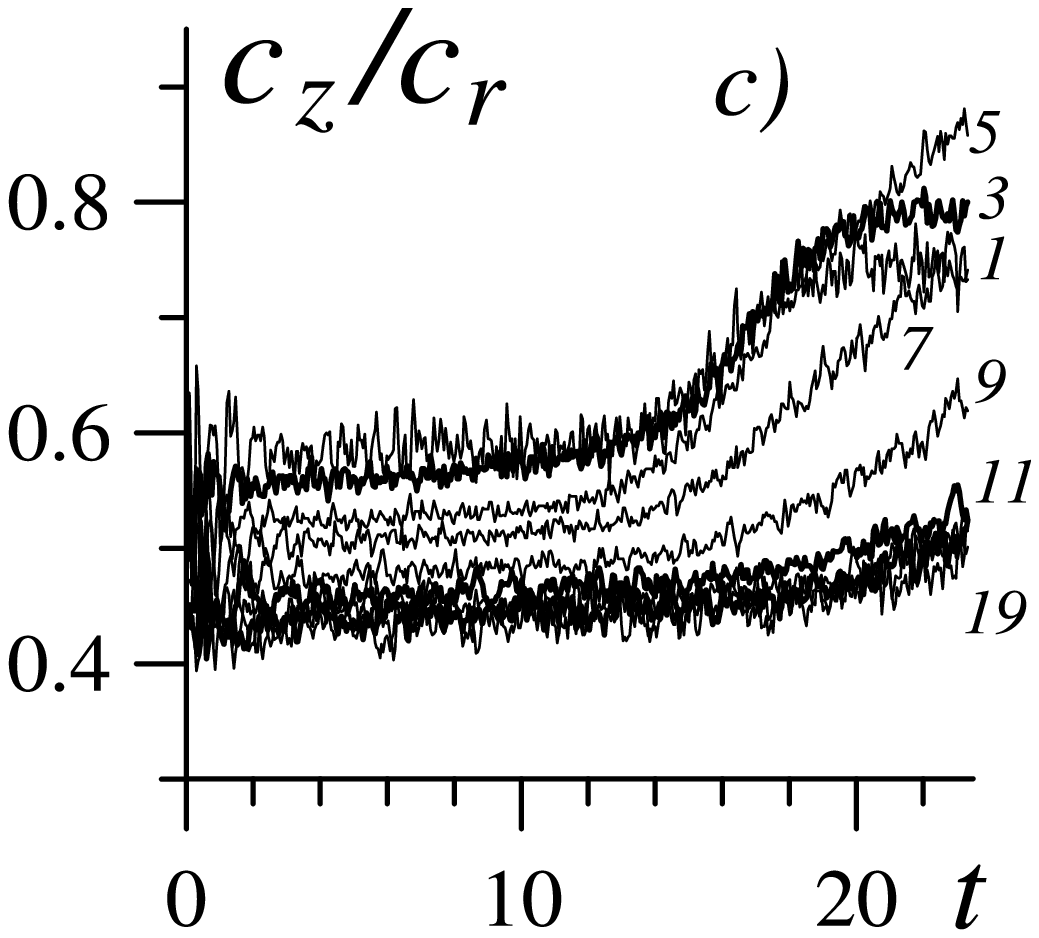}
\caption{ Results of the same model as in Fig.~\ref{fig1} but evaluated
with the help of PP code (see text).
}
\label{fig3}
\end{figure*}

We verified the numerical effects of our TREE-code by reproducing the
most important numerical experiments with a direct particle-particle
(PP hereafter) code, see description by \cite{AVH2003}. The analogue of
Fig.~\ref{fig1} that was obtained with the help of PP code that started
from the same initial conditions and was evaluated with only $5
\times 10^4$ particles is shown in Fig.~\ref{fig3}. The residuals
in Fig.~\ref{fig3} are greater than those in Fig.~\ref{fig1} due to
less number of particles. Nevertheless, all features from
Fig.~\ref{fig1} can be seen in Fig.~\ref{fig3}.

\begin{figure*}
\includegraphics[width=5.2cm]{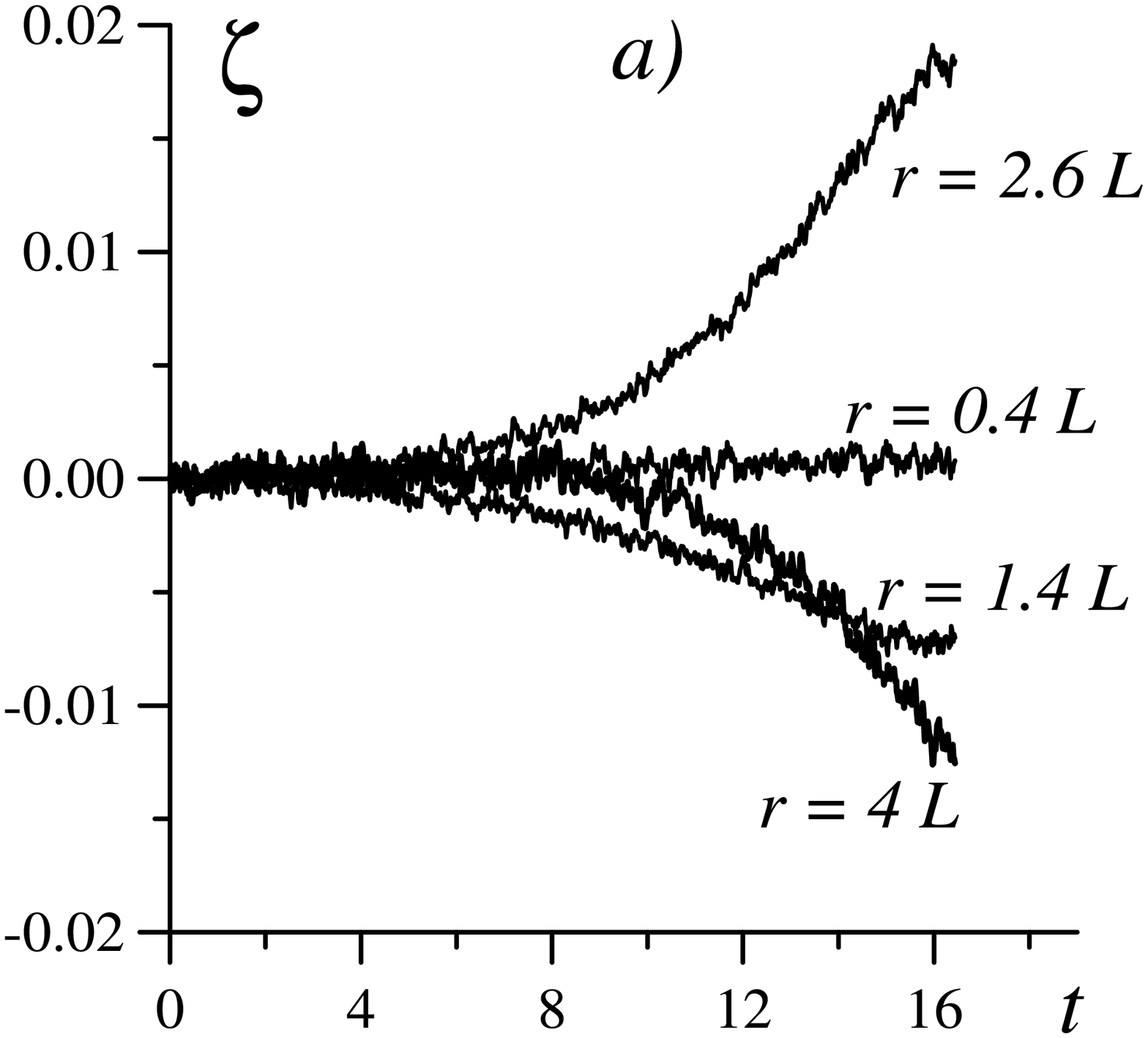}
\includegraphics[width=5.2cm]{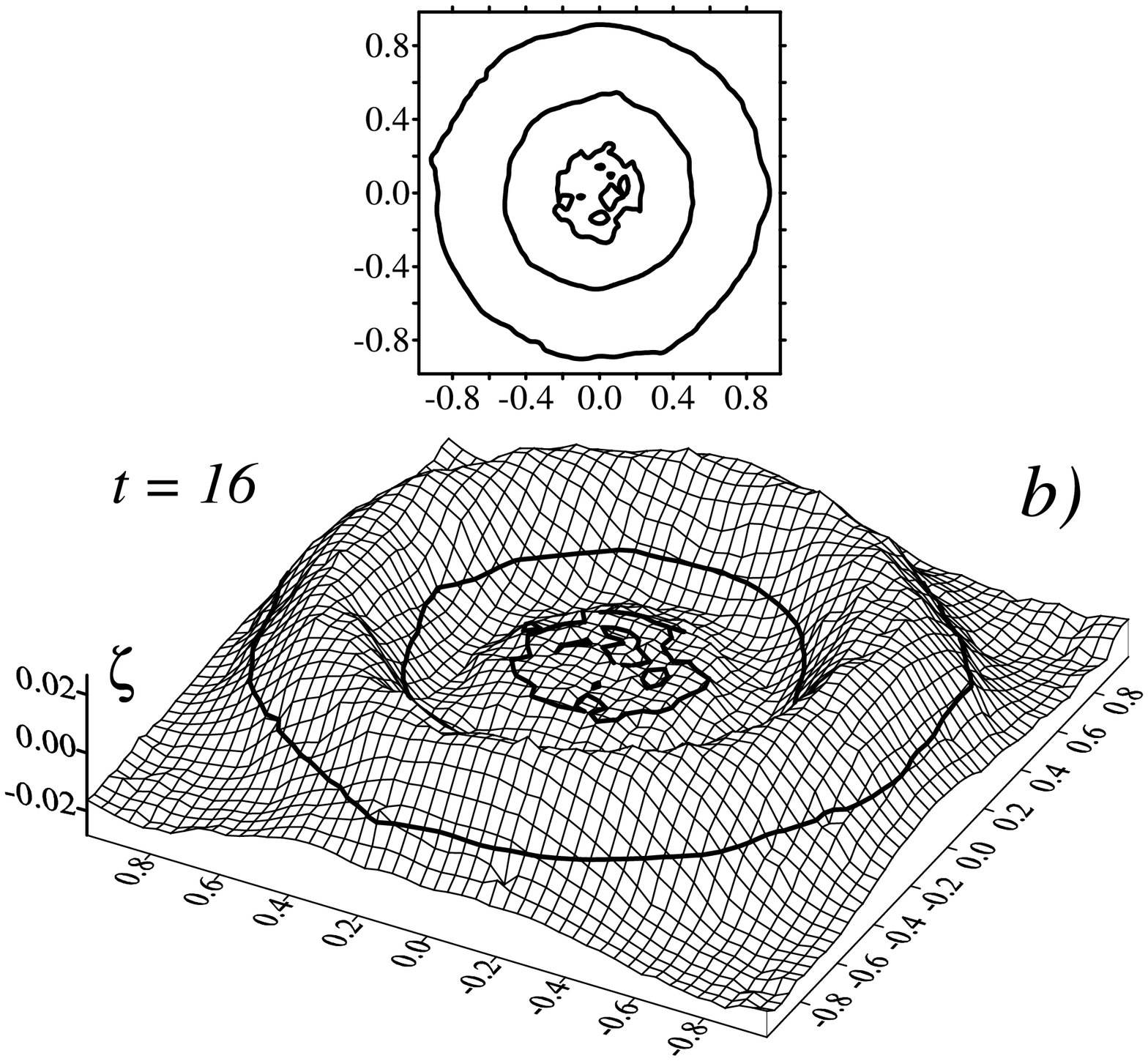}
\includegraphics[width=5.2cm]{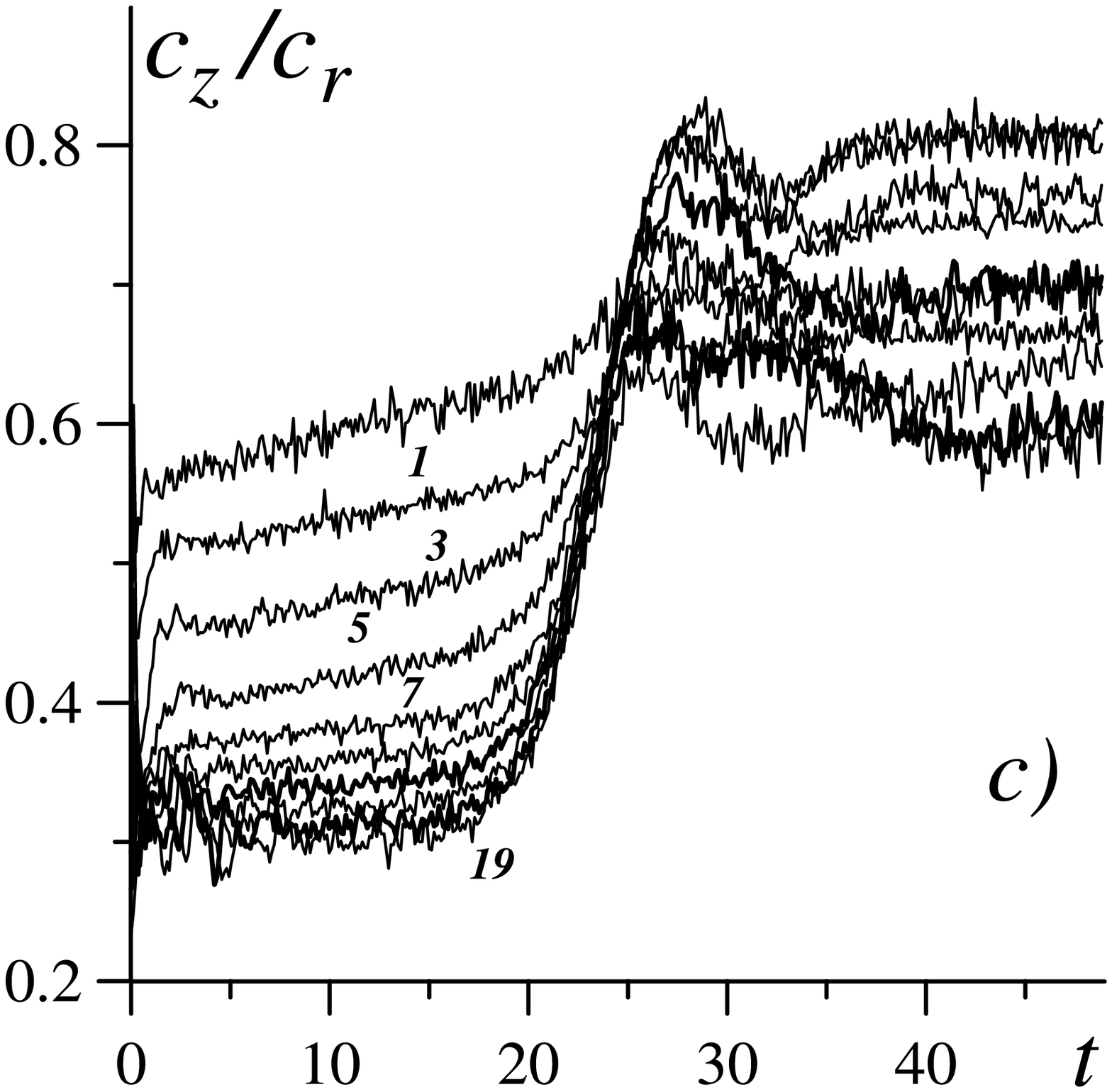}
\caption{ Initial conditions of this model with $\mu=4$ provide
the central parts of disc with stability, and the bending mode
$m=0$ develops itself in the outer regions ($r\gee L$). a) Initial
stages of evolution of $\zeta$ at different distances from the
centre. b) Distribution of $\zeta=0$ through the disc. The thick
line denotes the isoline $\zeta=0$. c) The heating of a model disc
without a halo at all, in presence of a bulge ($M_b=0.25\,M_d$, $b=0.2\,L$),
when the mode $m=0$ is being developed. The ratio $c_z/c_r$ does
not rise significantly in regions of the bulge. The notation is
kept the same as in Fig.~\ref{fig1}.
}
\label{fig4}
\end{figure*}

If the model disc is not massive enough to provide the gravitational
stability against the bar-mode and mass of the halo is low
($\mu\lee 1$), the main reason for heating the initially thin
cold disc is the axisymmetric bending mode ($m=0$) whereas the
modes $m=1,2$ do not emerge at all. Note that the mode $m=0$ may
emerge far from the centre ($r\gee L$) and develop there
without penetrating into the central part of the disc. It can happen
because of
two reasons: either the initial disc had $c_z/c_r \ge
\alpha_z^{crit}$ (see \S~3.4) and was stable in the central
regions whereas periphery of the disc is thin and unstable
(Fig.~\ref{fig4}), or the galaxy harbours a concentrated and massive bulge.
In such models the bending modes in the centre have low amplitude.
The presence of a bulge and/or a halo plays a stabilizing role. Hence,
having all other conditions equal, the discs of galaxies
possessing bulges have to look thinner. Note that since we did not
consider dynamically non-stationary bulges in our model, this our
conclusion requires further investigation.

\subsubsection{ Bending modes $m=1,2$ }

In the models with massive halo ($\mu\ga 2$), the
axisymmetric bending modes $m=1$ and $m=2$ may develop themselves
and heat up the stellar disc along its vertical direction. The surface
density of the disc remains axisymmetric.
The evolution
of initially thin disc for a case of a heavy halo $\mu=4$ is shown in
Fig.~\ref{fig5}.
In this case the bending mode $m=2$ is
developed initially in the inner disc regions ($r\la 2L$), see the
saddle-like features in Fig.~\ref{fig6}.
Fig.~\ref{fig6} shows the distribution of $\zeta$ in the disc
plane for certain moments between $t=0$ and $t=40$
(like in Fig.~\ref{fig5}).
The vertical heating and disc's flare
due to the mode $m=2$ are very modest in this case.
However, a non-linear
one-arm asymmetric mode $m=1$ emerges after $t\ga 3$ and the
vertical heating gets more significant. The third
stage of the heating starts at $t\ga 10$ when the
mode $m=0$ begins to dominate all over the disc. The growth
rate of the vertical dispersion $c_z$ is especially
high at this time: the disc scale height increases by
a factor of 2 -- 3. The example considered shows the
process of transformation between the bending modes
and transition from the asymmetric mode $m=2$ to the
axisymmetric one.
In the case of a low-massive spherical subsystem of
low mass, the favourable conditions for emerging
the global bending instability get worse and the disc warp does
not emerge in the disc for at least 20 rotation turns if
the initial $\alpha_z = c_z/c_r$ is high enough.

If the initial template was too far from the stability limit,
the disc "jumps over" the stability limit, and then
its final state essentially exceeds this limit. Thus finally produces
overstabilized and too thick disc. Since such stellar discs are not
obviously observed, we have to assume that the initial state of the
discs was not too far from the stability limit. Note that it agrees
with a gradual building of stellar discs from gas or via minor merging
when the initial
discs do not emerge instantly but are built for a continuous time
from small fragments, which adjust to the discs
stability limit.

\subsection{Bending of bars}

In the course of experiments with low massive halo ($\mu\lee 1.5$),
if the initial state of disc is gravitationally unstable, a bar
can emerge. The bar formation is accompanied by its warping
(as reported by \cite{Raha}).

Bending of the bar can be a result of global
instability of the bar-mode during its initial stage
when the bar forms in the initially thin cold disc. Fig.~\ref{fig7}
shows the bar bendings that increase the vertical velocity dispersion.
Amplitude of the bar warps decreases essentially when
the bar thickness increases. We stress
that once the bar has been formed, it stops the
further possible developing of the global bending
modes, and first of all it destroys the axisymmetric
instability with mode $m=0$.

The bar formation runs faster while the initial disc
is cold (i.e. for small values of Toomre's parameter $Q_T$) and as
a result, amplitude of bar's bending mode increases
(Fig.~\ref{fig7}). If the initial disc lays in a slightly subcritical
state ($c_r$ is just below the level that
provides the stability to the global bar-mode), the bar emerges very
slowly, and it is stable against the bending perturbations, i.e. the bending
of a bar does not necessarily emerge in the modelling.

\subsection{ The ratio $c_z/c_r$}

The key parameter responsible for stability of the stellar discs is
the ratio $\alpha_z=c_z/c_r$. To stabilize the global bending
instabilities in the case of low values of relative mass of the spherical
subsystem in a model, the value of $c_z/c_r$ has to be greater
than $0.3-0.37$, as it was figured out from linear analysis of
simple models (\cite{Polyachenko77, Araki-1986, Merritt}).

Let's consider a bulgeless model with a moderate halo $\mu=1$, see
Fig.~\ref{fig1} and ~\ref{fig2}. The initial (curve 1) and final
(curve 2) distributions of $\alpha_z(r)$ are shown in
Fig.~\ref{fig9}a. The vertical heating generated by the
axisymmetric bending mode is as strong that the averaged over the
disc ratio $\langle c_z/c_r\rangle=0.74$.

Once the relative mass of halo is chosen to be higher,
the ratio $c_z/c_r$ needed to marginally stabilize the bending
perturbations (we refer to it as the critical ratio $\alpha_z^{crit}$
hereafter) gets less. One can see the initial
(curve 1) and final (curve 2) distributions of
$\alpha_z(r)$ for the case of very massive halo with $\mu=4$
in Fig.~\ref{fig9}b. The curve 3 shows the initial distribution of
$\alpha_z(r)$, which is necessary to provide stability against the
global bending modes. At the same time, at the disc
periphery $0.27<c_z/c_r<0.37$.

A distinctive feature of the considered models at the threshold of
bending stability is the non-uniformity of $c_z/c_r$ along the
radius (see Fig.~\ref{fig9}). For the case of moderate halo
($\mu\la 1$) with the initial distribution of $c_r(r)$ that
suppresses the bar instability, the critical value of
$\alpha_z^{crit}$ is a descending monotonous function of $r$. Its
value ranges from 0.5 $\div$ 0.6 at the central regions to 0.3
$\div$ 0.4 at the periphery (see Fig.~\ref{fig3}).

\section{ Modelling of selected edge-on galaxies}

In order to compare our model predictions with observations, we
choose eight spiral edge-on galaxies. Structural parameters for
five of them: NGC~4738, UGC~6080, UGC~8286, UGC~9442 and UGC~9556, were
taken from \cite{dmbiz2004}. Superthin edge-on galaxy UGC~7321
was studied by \cite{Matthews1999, Matthews2000}. Radial
distributions of the stellar velocity dispersion are available
in addition to rotation curves and structural parameters for
two large and nearby galaxies NGC~891 and NGC~5170
(\cite{Kruit1981, Morrison1997, Bottema1991, Bottema1987}). All the
galaxies except NGC~891 and NGC~5170 indicate a negligible
presence of bulge, which simplifies the modelling.

We evaluate the following free parameters of the
model: $\mu = M_s / M_d$, the radial scale of the
halo $a_h$ and the
disc central surface density $\sigma_0$. We look for
the optimal agreement between the calculated and
observed disc thickness and rotation curve simultaneously.
Here $M_d$ is the galactic disc mass; $M_s = M_h + M_b$ is the mass
of the spherical component which comprises of the halo
and bulge in general case. Note that since almost all
our galaxies have no visible bulge, their spherical
component in the most cases means the galactic dark halo.

We assume that the density distribution in stellar discs
and bulges (if used) follows the brightness distribution.
It corresponds to assumption of constant mass-to-light ratio.

The structural parameters of the galaxies are figured out for the
Hubble constant $H_0=75$~km~s$^{-1}$~Mpc$^{-1}$. The name of
galaxy, adopted distance $D$, scale length $L$, observed mean scale heights
$\langle z_0\rangle$ and $\langle h_{z}\rangle$ (corresponding to
$\rm sech^2$ and $\exp$ luminosity distributions in the vertical
direction, respectively), stellar disc radius $R_{\max}$, and
reference to the source of the rotation curve are shown in Table \ref{Tabl1}.

\begin{table}
\begin{center}
\caption{Observational parameters of selected edge-on galaxies}
\label{Tabl1}
\smallskip
\begin{tabular}{lllccrc}
\hline
\hline
Name     & $D$   & $L$     & $\langle z_0\rangle$
&$\langle h_{z}\rangle$ & $R_{max}$ & RC \\
  & Mpc  & kpc   & kpc     & kpc     & kpc     &        \\
\hline
NGC~891  & 9.5   & 4.9     & 0.98    & 0.49    & 21.0 & 1 \\
NGC~4738 & 63.6  & 4.7     & 1.30     & 0.7    & 19.2 & 2 \\
NGC~5170 & 20.0   & 6.8     & 0.82    &  --    & 26.2 & 3 \\
UGC~6080 & 32.3  & 2.9     & 0.69    & 0.48    &  9.9 & 4\\
UGC~7321 & 10.0   & 2.1      & 0.17$^b$& 0.14$^c$& 8.15 & 5\\
UGC~8286 & 4.8  & 2.0     & 0.26     & 0.13    &  8.0 & 6\\
UGC~9422 & 45.6  & 3.5     & 0.80    & 0.51    & 14.6 & 4\\
UGC~9556 & 30.6  & 1.5 (3.6)$^a$& 0.51  & --   &  9.0 & 4\\
\hline
\end{tabular}
\smallskip
\end{center}
\vbox{\footnotesize
\noindent $^a$ two exponential discs are considered: HSB and LSB;
the latter one has larger scale length.

\noindent $^b$ the scale is given for the disc's periphery in case
$\textrm{sech}(z/z_{ch})$.

\noindent $^c$ the scale is given for the disc's centre.

\noindent Here $D$ is the distance to the galaxy, $L$ is the
exponential disc scale length, $\langle z_0\rangle$ is the scale
height for the $\textrm{sech}^2$ shape of the vertical luminosity
profile, $\langle h_z\rangle$ is the scale height for the case of
exponential vertical profile, $R_{\max}$ is the disc cut-off
radius. References to the rotation curves: 1 --
\cite{Sancisi1979, Broelis1991}, 2 -- \cite{Giovanelli-1997}, 3
-- \cite{Bottema1987} 4 -- \cite{IDK1991}, 5 --
\cite{Matthews1999}, 6 -- \cite{Bottema1986}. }
\end{table}

\subsection{Notes on individual objects}
\label{sec4.1}

Below we present some results and notices on modelling of the
galaxies.
All masses of galactic components given below are estimated within
the limits of $R_{\max}$ from Table 1.

\subsubsection{UGC~9422}

Fig.~\ref{fig10} shows the rotation curves $V^{obs}$ together with
the model circular velocity curves $V_c$. $V^{obs}$ are shown for
$\mu = 0.35$ and $\mu = 0.6$ (dashed and dash-dotted curves,
respectively). The "edge-on" values of $V_c$ are plotted for the
same $\mu = 0.35$ and $\mu = 0.6$ (dotted and solid curves,
respectively). The model radial distributions of $z_0$ are shown
in the top right panel in Fig.~\ref{fig10}. Model with low massive
halo ($\mu = 0.4$, the solid curve) produces too thick stellar
discs. It rules out the maximum disc model in which the central
surface density $\sigma_0=1440\,M_\odot/$pc$^2$ and the disc
contributes 94\% into the circular velocity ($V_{c}^{disc}/V_{c}$)
at $r$=2.2L. The acceptable agreement with the observed disc
thickness can be achieved at $\mu \simeq 0.6$ (the dotted curve).
Haloes with $\mu \ga 1$ make the disc to be too thin (the model
with $\mu=1.6$ is shown by the dashed curve in the corresponding
panel of Fig.~\ref{fig10}). The best-fitting model has parameters:
$M_d=8.6\cdot 10^{10}\,M_\odot$, $M_h=5.2\cdot 10^{10}\,M_\odot$,
$\sigma_0=1200\,M_\odot/$pc$^2$ and
$V_{c,d}/V_{c}|_{r=2.2L}=0.87$.

\subsubsection{UGC~6080}

The $V_c$ and $V^{obs}$ in the left panels of Fig.~\ref{fig10}
for UGC~6080 are designated by the solid and dashed curves,
respectively. The model disc thickness is shown in the right panel
for $\mu=0.36$ (solid curve), $\mu=0.57$ (dotted curve) and
$\mu=0.9$ (dashed curve).
The stellar disc is thicker than it is observed for all models with
$\mu<0.6$.  The best-fitting model gives the upper limit for the
disc mass $M_d=5.3\cdot 10^{10}M_\odot $ and $\sigma_0=1200\,
M_\odot/pc^2$ for its central surface density.

\subsubsection{UGC~8286}

Left panel for UGC~8286 shows only $V^{obs}$ for two models:
$\mu=0.7$ (solid curve) and $\mu=1.6$ (dotted curve).
Right panels show radial distribution of observed disc thickness
for the same cases (solid and dotted curves, respectively).
It is a good illustration that a
rotation curve can be successfully explained with a wide
range of $\mu$ whereas the disc thickness (right panel in Fig.~\ref{fig10})
helps to choose the proper model.
The best-fitting disc central surface density is
the lowest in our sample that reveals the LSB nature of this galaxy.

\subsubsection{UGC~9556}

The radial photometric profile of UGC~9556 cannot be
described by a single exponent. We find that It is better fit by
two exponents with the scale lengths 1.5 and 3.6 kpc (\cite{dmbiz2004}).
The notation in the left panel is kept the same as for UGC~6080.
In the right panel, we show the model thickness for $\mu=0.9$ (dotted curve)
and $\mu=1.9$ (solid curve).
The best-fitting model with $\mu \approx 0.9$
produces a bar which can be responsible for the deviations
from the exponential law in the inner regions of the model disc.

\subsubsection{UGC~7321}

Thin stellar disc of UGC~7321 and a lack of bulge indicates that
the galaxy owns a massive dark halo. The notation in the left
panel is kept the same as for UGC~6080. In the right panel we show
the model thickness for $\mu=3.4$ (dotted curve) and $\mu=1.8$
(solid curve). The best-fitting model with $\mu=3.4$ has
$M_d=0.53\cdot 10^{10}\,M_\odot$ and $M_h=1.8\cdot
10^{10}\,M_\odot$. The stellar disc contributes $\simeq 60$~\%
into the circular velocity at $r=2.2L$. The central surface
density in this case is only $\sigma_0=220\,M_\odot/$pc$^2$. This
highlights the low surface brightness nature of the disc inferred
by \cite{Matthews1999}. We investigate a case of the bar formation
in this galaxy, according to the hypothesis by \cite{Pohlen2003}
and make the conclusion that it cannot exist in UGC~7321,
according to our modelling.

\subsubsection{NGC~4738}

This galaxy has a bulge according to the photometric data, but its
luminosity is less than 3\% of the disc's that enables us to
consider bulgeless models for NGC~4738. In the right panel we show
the model thickness for $\mu=0.6$ (dotted curve) and $\mu=0.3$
(solid curve). The halo with intermediate mass $\mu \ga 0.5$ is
suitable for this galaxy. The models with less massive halo
($\mu=0.3$) produce too thick discs for this galaxy.

\subsubsection{NGC~5170}

An extended bulge is required to match observed RC for this galaxy.
On the other hand, the photometry indicates a presence of very
small bulge in this galaxy. The observed shape of RC with rather low
resolution so a disagreement with the model RCs is not unexpected
at $r\la 2$ kpc.
Fig.~\ref{fig11} shows the RC and radial distributions of velocity
dispersions calculated for the maximum disc model and for the
model with $\mu = 1.6$. The latter case provides a better agreement
with observational data for both RC and velocity dispersion.
Moreover, the maximum disc model requires the main disc thickness to be
1.1 kpc whereas the observations suggest 0.82 kpc. The
model with $\mu = 1.6$ produces the projected mean scale height 0.8
kpc. In general, the agreement between RC and both velocity
dispersion (Fig.~\ref{fig11}c) and stellar disc thickness is matched
when $\mu\ga 1.5$.

\subsubsection{NGC~891}

This galaxy has a budge (\cite{Bottema1991, Morrison1997}),
and we took it into account evaluating our numerical models.
Published values of the vertical exponential scale height for old
stellar disc varies from 400~pc to
650~pc (\cite{Kruit1981, Bahcall1985, Shaw, Morrison1997, Xilouris-1999}).
We assume the set of structural parameters inferred for NGC~891
by \cite{Kruit1981}.

The observed and model radial distributions of the stellar
velocity dispersion (\cite{Bottema1991}), stellar disc thickness
and rotation curves are shown in Fig.~\ref{fig12}.
The stellar velocity dispersion in the centre of the galaxy
reaches 160~km/s (\cite{Bottema1991}) for stars
in the bulge, and the radial component of the disc velocity dispersion
is $c_r\le 120$~km/s. Existence of a
molecular ring in the galaxy at $r\simeq 3.5 $~kpc enables us to
assume that the galaxy harbours a bar (\cite{N891-Sofue}).

Once the radial distribution of the stellar velocity dispersion is
taken into account together with condition of marginally stable
stellar disc, successful model should have $\mu>1.5$. Note that the use
of an alternative RC from \cite{AVH2001} gives $\mu=1.8$ with
other similar parameters.

Fig.~\ref{fig12}b shows the radial distribution of the disc
exponential vertical scale height for the set of $M_h/M_d$ = 0.4,
0.7 and 1.7. The stellar disc thickness indicates that the dark
and luminous masses are roughly equal to each other within
$R_{\max}$. The value of $\mu$ inferred from the disc thickness is
slightly less than that estimated from the velocity dispersion
probably due to additional factors that increase the disc
thickness in addition to the bending instability.

The disc mass of $1.5\cdot 10^{11}\,M_\odot$ estimated
with the shape of rotation curve (\cite{Bottema1991})
corresponds to a very high value of
mass-to-light ratio 13.5 $\pm$ 4.
Our method yields $5\div 10$ for this ratio.
This introduces an argument against the the maximum disc in NGC~891.

Finally, we consider a case of two stellar discs of different thickness
in the galaxy. This assumption was made by \cite{Morrison1997}.
We found that such an assumption almost does not change the final
best-fitting value of $\mu$.

\subsection{Relative thickness of stellar disc versus relative mass of
spherical subsystem}

A tight connection between the stellar disc thickness $z_0/L$ and
relative mass of spherical subsystem follows from our modelling,
see Fig.~\ref{fig14}. Each circle in the Fig.~\ref{fig14} corresponds
to a model galaxy with different parameters of the disc and spherical
subsystem. The figure includes our modelling of the objects studied
by \cite{AVH20012,Zasov1991} in addition to results of the
present paper. The upper curve corresponds to mostly bulgeless
models. The lower one designates the loci of models with noticeable
and compact bulges. As it can be seen in Fig.~\ref{fig14}, the presence of
a bulge introduces a little scatter to the $<z_0>/L \div M_s/M_d$ dependence
that can be easily parameterized in dependence of the bulge mass.
This relation can be used to estimate $M_s/M_d$ if $z_0/L$ is known
from observations.

\subsection{Results and discussion}

The parameters of best-fitting models for our galaxies are
summarized in Table~2.
In addition to the evaluated model parameters, we estimate the
dark-to-luminous mass ratio $M_{dark}/M_{lum}$ (it equals
to $\mu$ if no bulge was assumed in the model.
In the case of bulge we derive the ratio $M_{dark}/M_{lum}$
as $M_h/(M_d + M_b)$, where $M_b$
denotes mass of the bulge. As one can see from the table,
the fractions of dark and luminous matter in more than a half objects
from our sample are roughly
the same. The exceptions are the thin galaxies NGC~5170, UGC~7321 and
UGC~8286 which have massive dark haloes and low surface brightness
discs (low $\sigma_0$ in our modelling). The same may be referred to
the external low surface brightness disc of UGC~9556.

In Fig.~\ref{fig13} we compare the relative disc thickness
($z_0/L$) and its central surface density (here we consider
only thin external LSB disc in UGC~9556). The figure shows a
good correlation between the disc relative thickness and its central
surface density. Our sample is small but much more
data from photometry can be incorporated.
Thus, the disc central surface brightness was compared with their 
thickness by \cite{dmbiz2002,dmbiz2004,dmbiz2009}.
A good observational correlation was found: the thicker the disc, the
greater its central surface brightness. The latter parameter reveals the
central surface density of disc and
corresponds to $\sigma_0$ in Fig.~\ref{fig13}.
It means that the correlation in Fig.~\ref{fig13} comes
from two independent methods: structural studies and
numerical modelling.

The relative thickness of the disc is a crucial parameter in our
modelling, and is relevant to the relative mass of the
spherical component.
The relative thickness may span quite a wide range
from 3 to 12 (see \cite{Kruit1982,
Grijs-disser, Grijs-Kruit-1996, dmbiz2002, dmbiz2004, dmbiz2009}).
According to our conclusions it
means that the relative mass of the spherical subsystem
varies significantly from galaxy to galaxy.

All these values are dimensionless
whereas the central surface density of stellar disc
depends on the amplitude of rotation curve which comes
from observations. Hence, our model could reproduce a
rather thick low surface brightness disc (say, without a
spherical subsystem at all) or, in a contrary, a very dense
stellar disc surrounded by massive halo which makes
it to be very thin. As it can be seen from Fig.~\ref{fig13}
and \cite{dmbiz2002}, such kind of galaxies were not found
among both normal and low surface brightness spirals. It
suggests that some factors beyond our model assumptions regulate
the connection between the disc surface density and dark halo masses.
Thus can be a result of evolution as well as a consequence of
dark-luminous matter interaction on the stages of galaxies formation.

\begin{table*}
\begin{center}
\caption{Inferred parameters of the spherical and disc components
in the considered edge-on galaxies}
\label{Tabl2}
\smallskip
\begin{tabular}{lcccccccc}
\hline
Name & $a_h$ & $M_h$ & $M_d$ & $\sigma_0^{disc}$ & $a_b$ & $M_b$ & $\mu$ & $M_{dark}/M_{lum}$ \\
     & kpc   & 10$^{10}~ M_{\odot}$ & 10$^{10}~ M_{\odot}$ & $\frac{M_{\odot}}{pc^2}$ & kpc & 10$^{10}~ M_{\odot}$  && \\
\hline
NGC 891  &5.4  & 7.13 & 10 &  715 & 1.09& 1.86& 0.9 (1.7) & 0.7 (1.2) \\
NGC 4738 &10.4 & 5.5 & 11 &  870 & --  & --  & 0.5   &  0.5  \\
NGC 5170 &8.67 & 19.2 & 12.8 &  491 & 2.43& 1.15& 1.6   &  1.4 \\
UGC 6080 &3.53 & 3.01 & 5.3  & 1230 & --  & --  & 0.6  &  0.6 \\
UGC 7321 &2.16 & 1.8 & 0.53 &  220 & --  & --  & 3.4   &  3.4  \\
UGC 8286 &3.1 &0.76  &0.47  & 203 & -- & -- & 1.6 & 1.6  \\
UGC 9422 &3.8  & 5.2  & 8.6  & 1200 & --  & --  & 0.6   &  0.6  \\
UGC 9556 &3.7  & 1.29 & 0.57 (0.81) & 715 (139) & --  & --  & 1.1  &  1.1 \\
\hline
\end{tabular}
\smallskip
\end{center}
\vbox{\footnotesize
Name of galaxy, scale length of dark halo, mass of halo,
mass of disc, disc central surface density, scale
length of bulge (if a bulge was included), mass of bulge,
spherical-to-disc mass ratio $\mu$, and dark-to-luminous mass ratio.

\noindent The numbers in parentheses for UGC~9556 correspond to
its outer disc. For NGC~891, the estimates of $\mu$
and $M_{dark}/M_{lum}$ that were obtained from the observed radial velocity
dispersion are given in parentheses.
}
\end{table*}

Additional factors as density waves, tidal perturbations or GMO
can also be developed in stellar discs and amplify the vertical
motions in disc. Nevertheless, developing the bending
perturbations is the most powerful mechanism heating up the
discs and our model results should not underestimate $\mu$ very much.
As is was shown by \cite{ZHT04}, our approach works well for the
stellar discs of spiral galaxies Sa-Sd.

Structural parameters of considered galaxies were obtained for mostly
old stellar population (\cite{dmbiz2004}), therefore we do not expect
large variations of mass-to-light ratio in the vertical direction.

The galactic dust layer in edge-on galaxies creates a problem of reliability
of structural parameters. Nevertheless, the parameters for our
galaxies were obtained using off-plane regions (see full discussion
in \cite{dmbiz2004} and \cite{dmbiz2002}). It should significantly
decrease the influence of dust to results of our modelling.

Realistic simulations should include "live" halo rather that that
approximated by a fixed potential. For the case of the "live" halo
our N-body simulations illustrate an extreme case and the ratio
$M_s / M_d$ found in our models shows the lower limit for the
halo-to-disc mass ratio. We found an ability to run our model upgraded with
a "live" halo for one of our galaxies, UGC~8286. Figure~\ref{fig15}
is the same as the third upper and right panel in Figure~\ref{fig10}
(i.e. its part that concerns the disc thickness in UGC~8286)
updated with more realistic N-body
simulations. The asterisks (upper curve), diamonds (next lower), and
triangles designate the models with N$_{halo}$  = 2$\cdot$10$^5$,
4$\cdot$10$^5$ and 6$\cdot$10$^5$, respectively, where
N$_{halo}$ is the number of bodies in the dark halo in our
N-body simulations. Note that the curves
for N$_{halo}$  = 4$\cdot$10$^5$ and 6$\cdot$10$^5$ are almost identical.
As we can see, the more
realistic case of the three considered haloes suggests a more massive
spherical component than that for the fixed potential, but the difference is
of the order of 10\%. Thus, we encourage to consider our halo-to-disc mass
ratio given in Table~\ref{Tabl2} as a lower limit for the real value
of $M_{dark}/M_{lum}$ that is nevertheless not far away from it.

\section{Conclusions}

\noindent 1. Development of the bending instabilities in stellar
galactic discs is studied with the help of N-body numerical
simulations. The axisymmetric bending mode ($m=0$) is found to be
the strongest factor which may heat up the disc in the
$z$-direction. If no bar is developed, the bending modes $m=1$
and $m=2$ are not significant contributors to the process of
the disc thickening.
The most
significant growth of the disc thickness occurs at the initial
non-linear stage of the bending formation. Once the bending was
destroyed, the vertical heating gets less effective. The lifetime
of the bending mode $m=0$ is higher for the larger
relative mass of the spherical subsystem $\mu$. We show that
initially very thin discs increase their thickness much more rapidly
than those started from a marginally subcritical state. As a result of
the global bending instability development, the final values of
the vertical velocity dispersion $c_z(r)$ and vertical scale
height $z_0(r)$ can be larger than those required for the
fulfillment of the disc stability condition.

\noindent 2. The critical value of the ratio
$\alpha_z^{crit}=c_z/c_r$ is a function of the
spherical subsystem parameters. At the threshold of stability, the
value of $\alpha_z^{crit}(r)$ decreases with the distance to the
centre. The value of $\alpha_z^{crit}$ can be twice as less at the
periphery in the comparison with $\alpha_z^{crit}$ at the centre.
The radial
trend of $(c_z/c_r)^{crit}$ can be approximated as $c_z/c_r\propto
\exp(-r/L_\alpha)$ with very long scale length $L_\alpha\simeq (5-6) \cdot L$.

\noindent 3. The average relative vertical disc scale height
$\langle z_0\rangle/L$ decreases when the relative mass of the
halo $\mu$ increases. This can be utilized to estimate the mass
of the spherical subsystem for edge-on galaxies.

\noindent 4. In the frames of our approach,
we conduct N-body modelling of eight edge-on
galaxies and compare results with published rotation curves and surface
photometry data.
For the cases of NGC~5170 and NGC~891 we additionally incorporate
published data on the stellar radial velocity dispersion. The relative
mass of spherical subsystem (the pure dark halo in most cases) is estimated
for all the galaxies. We conclude that the mass of
the dark halo within the optical limits in a half of our galaxies is of the
order of their disc's mass.
The exceptions are a thin LSB galaxy UGC~7321 whose the dark halo
encompasses about 2/3 of the total overall mass, and two more
galaxies NGC~5170 and UGC~8286 with thin discs.

\noindent 5. The central surface density of stellar discs in the
galaxies of our sample correlates well with the relative
thickness of the discs. This correlation may be utilized to estimate
the thickness of stellar discs in non edge-on galaxies.

\section*{Acknowledgments} 
Authors wish to thank prof. A.V. Zasov for
fruitful discussions and the anonymous referee for valuable comments. The
research was partly supported by grants RFBR 09-02-97021 and by the Federal
Target Program "Scientific and scientific-pedagogical personnel of the
innovative Russia" 2009-21(7). The computations were partly conducted at the
supercomputer SKIF MGU Chebyshev (Moscow State University).


\clearpage


\clearpage


\clearpage


\clearpage


\clearpage
\begin{figure*}
\includegraphics[height=21cm]{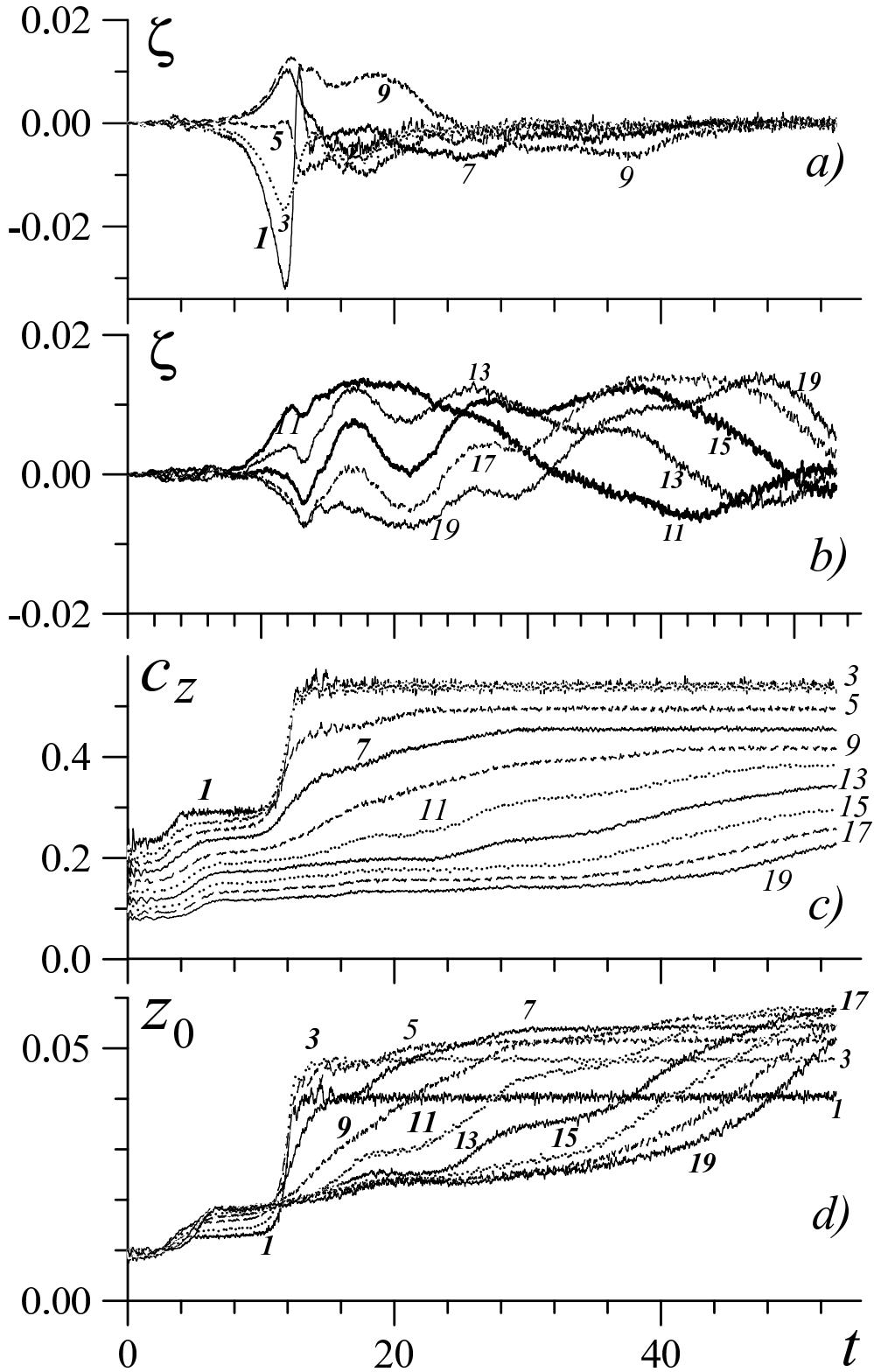}
\caption{
The same as in Fig.~\ref{fig1} but for the model with $\mu = 4$.
Distributions of $\zeta$ for the internal and
external regions are shown in separated two top panels.
}
\label{fig5}
\end{figure*}


\begin{figure*}
\setcounter{figure}{5}
\includegraphics[height=21cm]{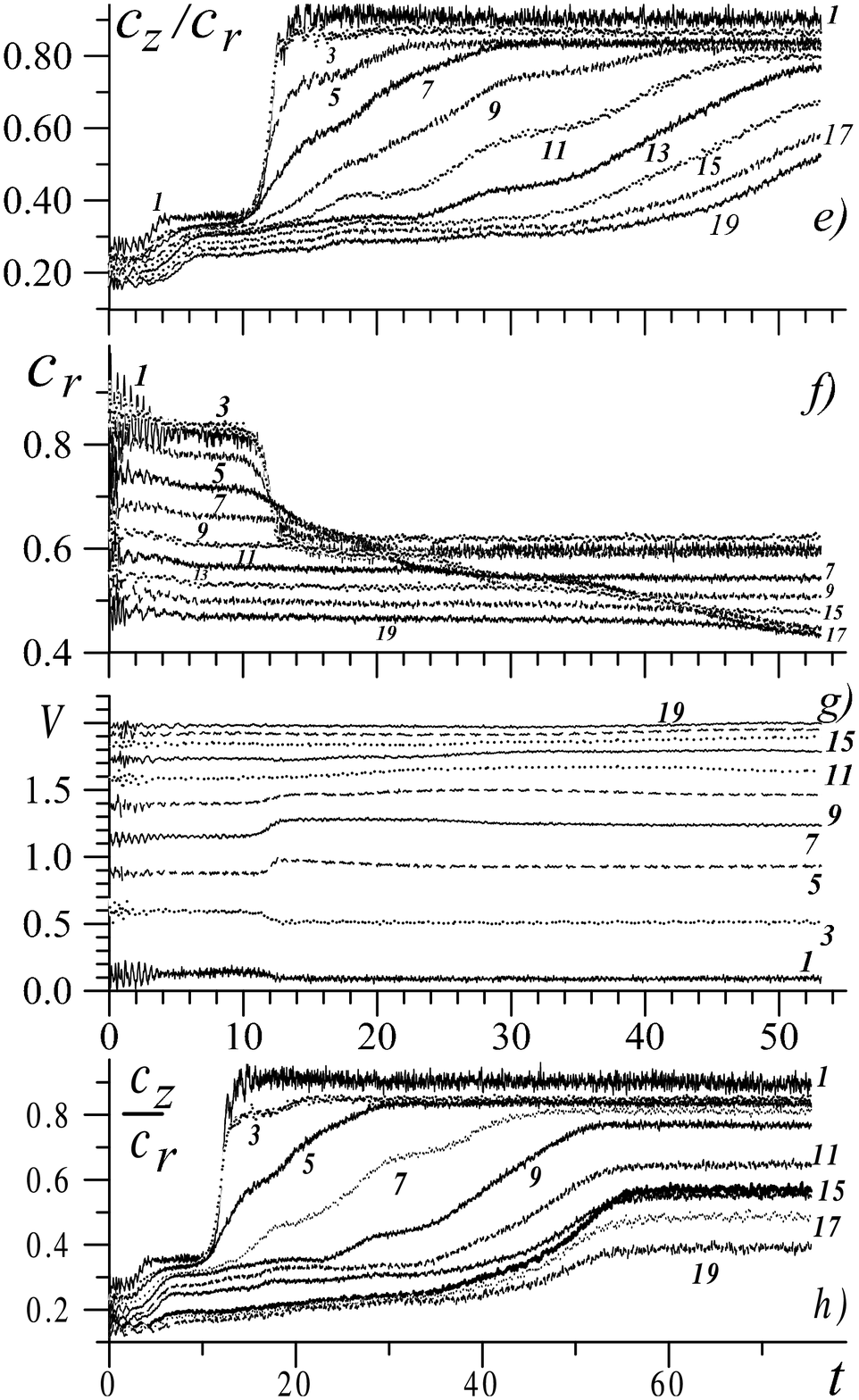}

\end{figure*}


\clearpage
\begin{figure}
\includegraphics[width=14cm]{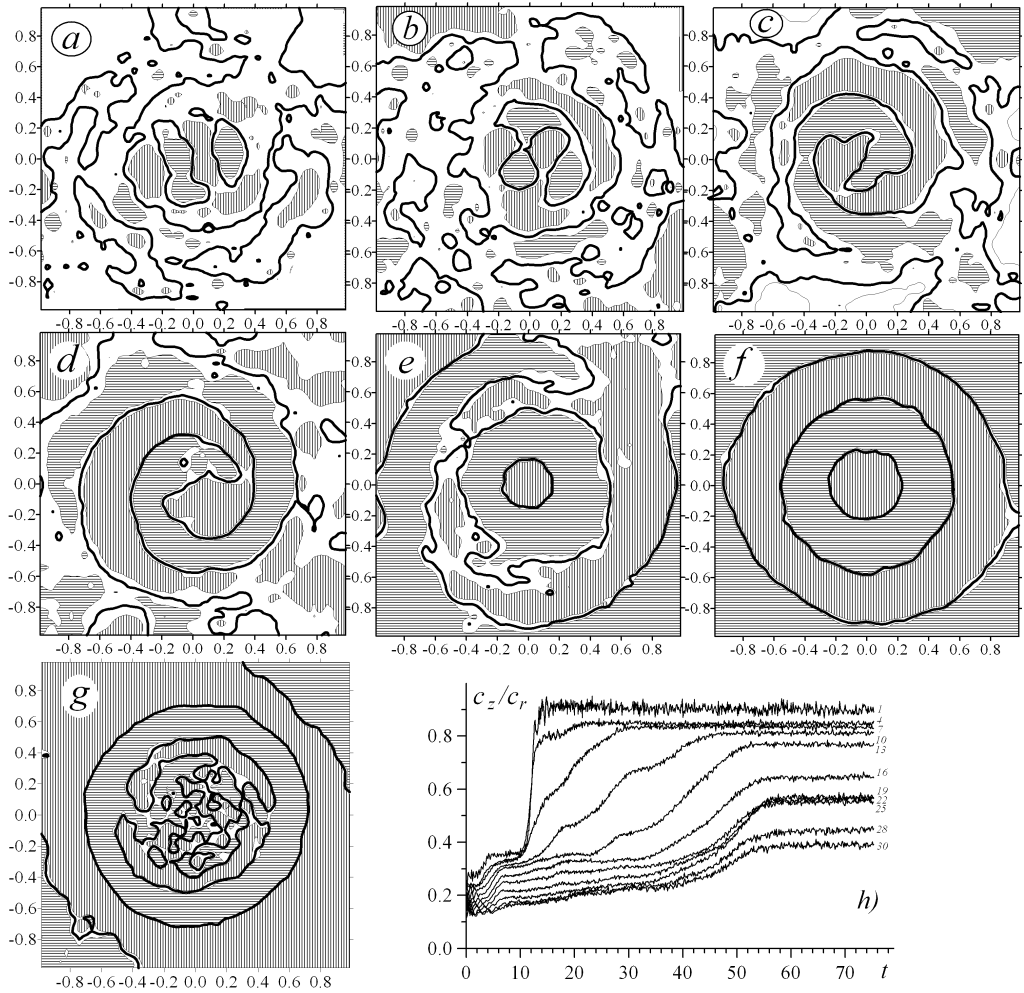}
\caption{
Distribution of $\zeta(x,y)$
in the disc plane at different moments of evolution:
a) $t=2.7$ -- the mode $m=2$ is developing in the central region,
b) $t=3.1$ -- $m=2 \rightarrow m=1$ reorganization,
c) $t=3.6$, d) $t=4.2$ -- the well-developed one-arm mode $m=1$ is seen,
e) $t=10.1$ -- $m=1 \rightarrow m=0$ reorganization,
d) $t=16.4$ -- the well-developed one-arm mode $m=0$ emerges,
e) $t=51$ -- all perturbations in the centre $r < L=0.25$ have almost
disappeared.
The axisymmetric perturbations can be seen in the intermediate
regions whereas the mode $m=2$ dominates at the periphery (f, g).
h) The time dependence of $c_z/c_r$ at different radii.
}
\label{fig6}
\end{figure}


\clearpage

\begin{figure*}
\includegraphics[height=9cm]{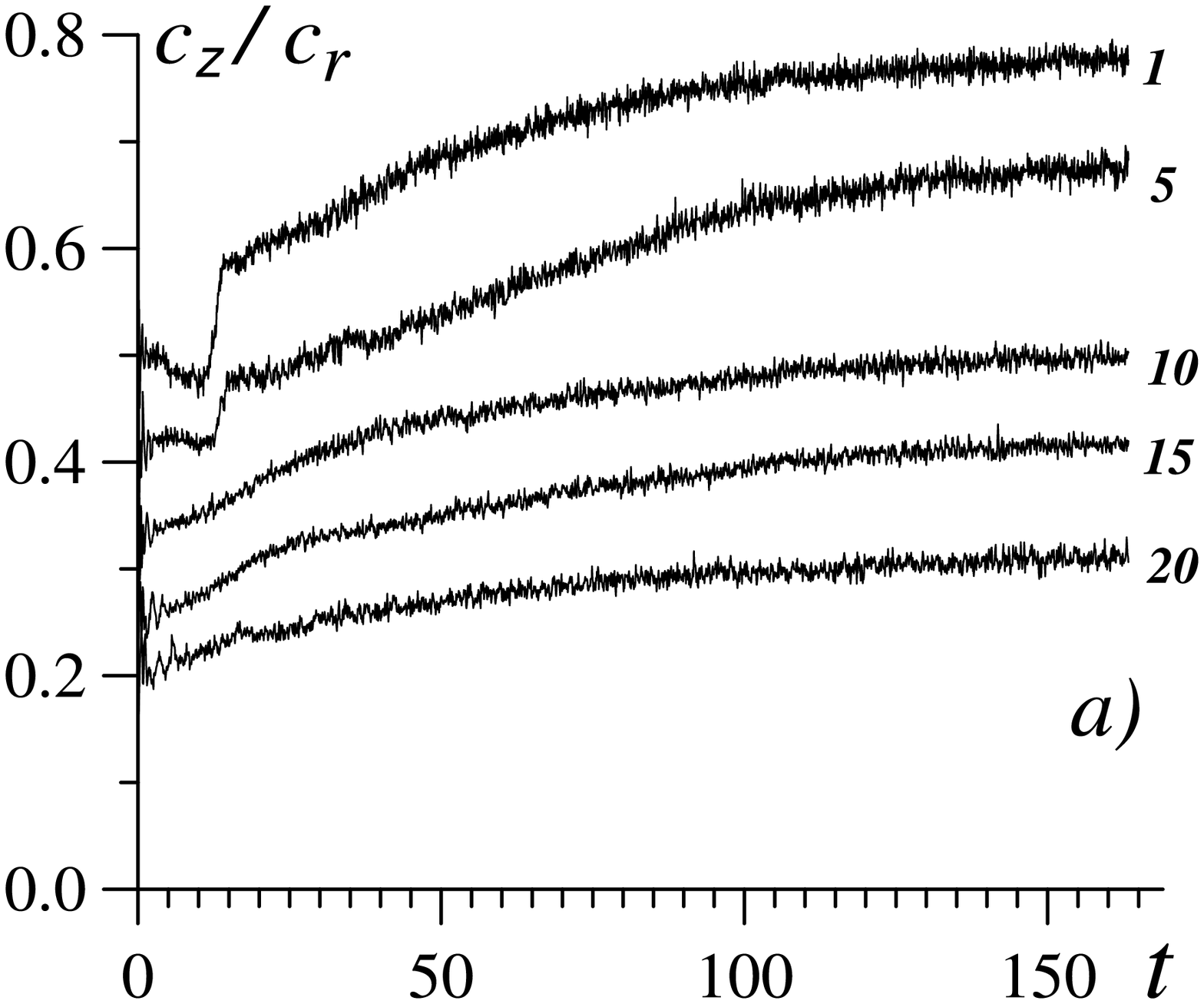}
\includegraphics[height=10cm]{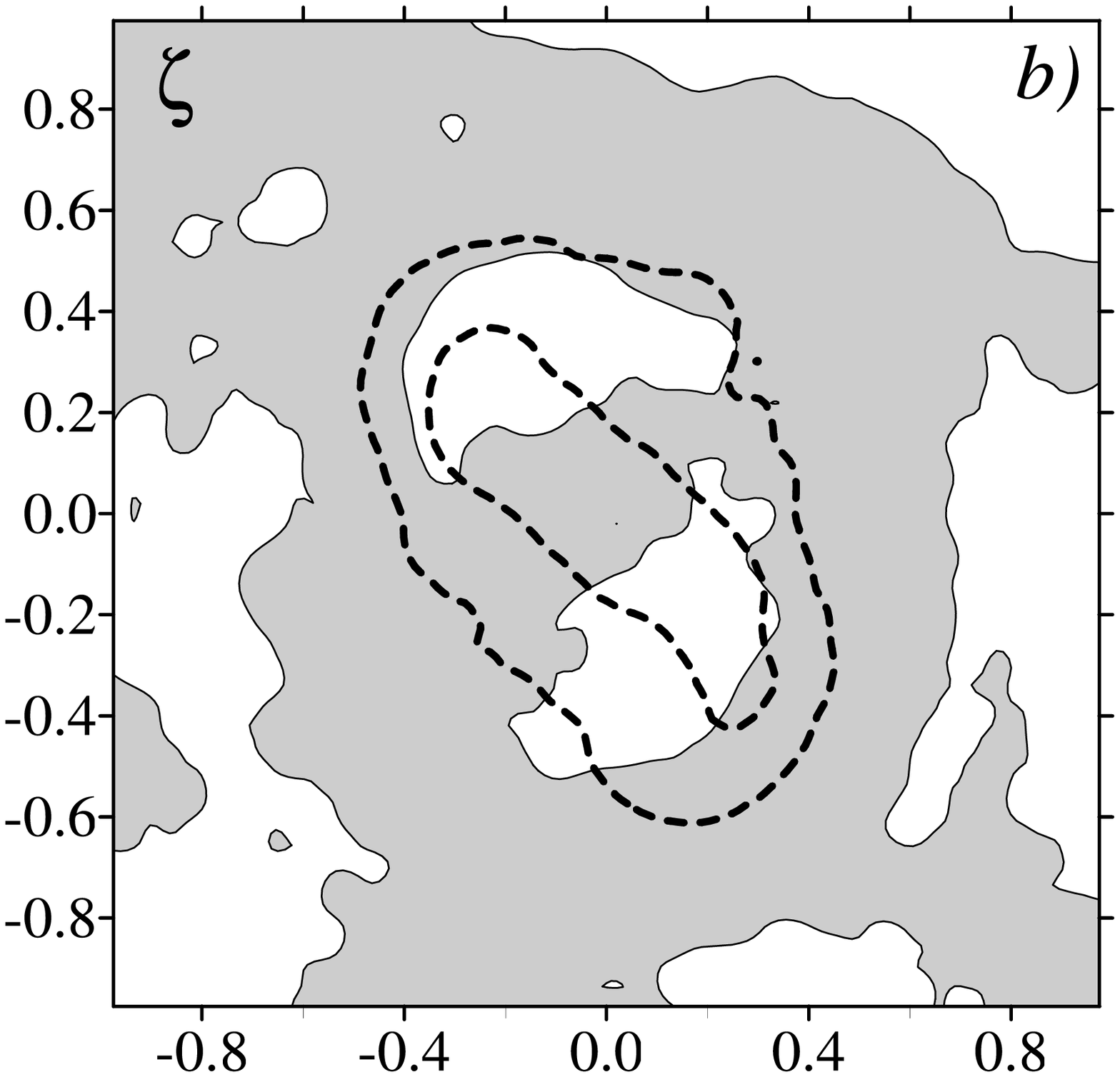}

\caption{ Formation of bar.
a) Evolution of the ratio $c_z/c_r$ at different radii in the model
 with bar.
b) Distribution of $\zeta$ in the plane $x-y$ reveals structure
of the bending mode at $t=15$. The thin solid line designates
$\zeta = 0$. The dotted lines show isodenses of the bar. Warping of
the bar is the reason for rapid vertical heating in the central
region (lines 1 and 5 at the range of $t=10\div 20$), see text.
}
\label{fig7}
\end{figure*}

%
%
%


\clearpage
\begin{figure*}
\includegraphics[height=6cm]{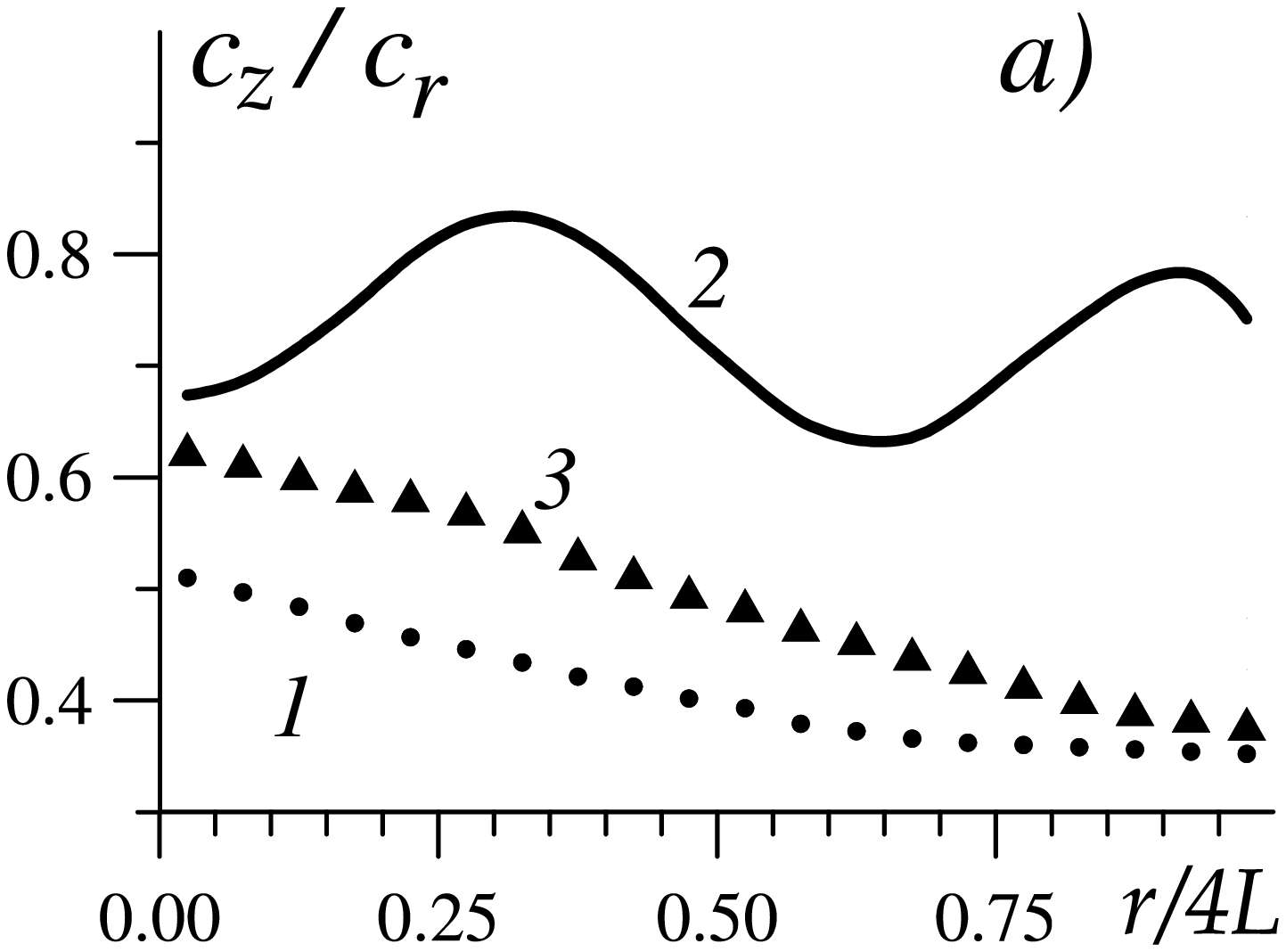}
\includegraphics[height=6cm]{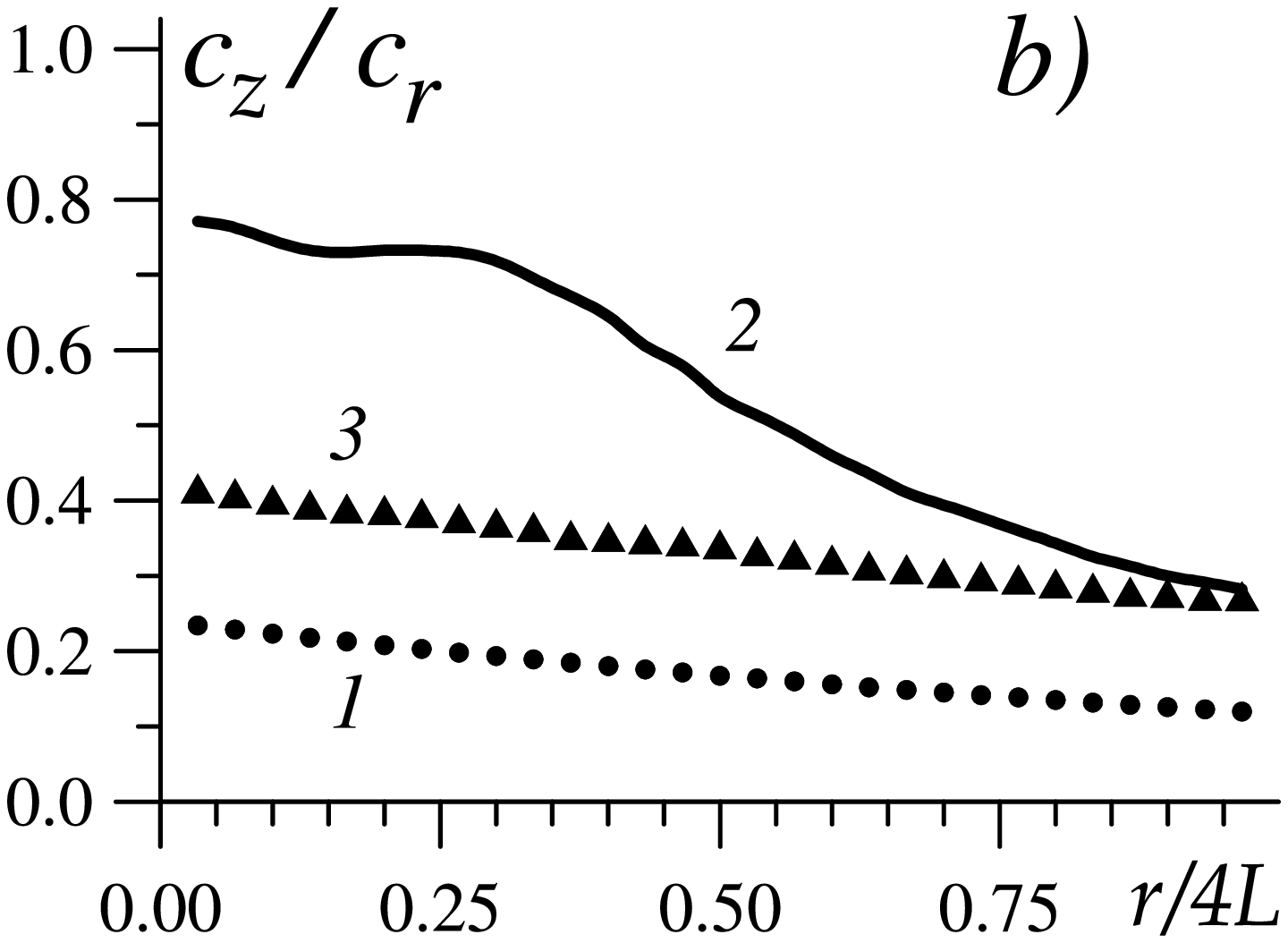}
\caption{ Radial distributions of $\alpha_z=c_z/c_r$ for a)
$\mu=1$ (see Fig.~\ref{fig1}) and b) $\mu=4$ (see
Fig.~\ref{fig5}). The dotted curves (1) are for the initial distribution,
solid lines (2) designate the final radial distributions and
triangles (3) denote the critical levels of $\alpha_z$.
} \label{fig9}
\end{figure*}


\clearpage
\begin{figure*}
\includegraphics[width=16cm]{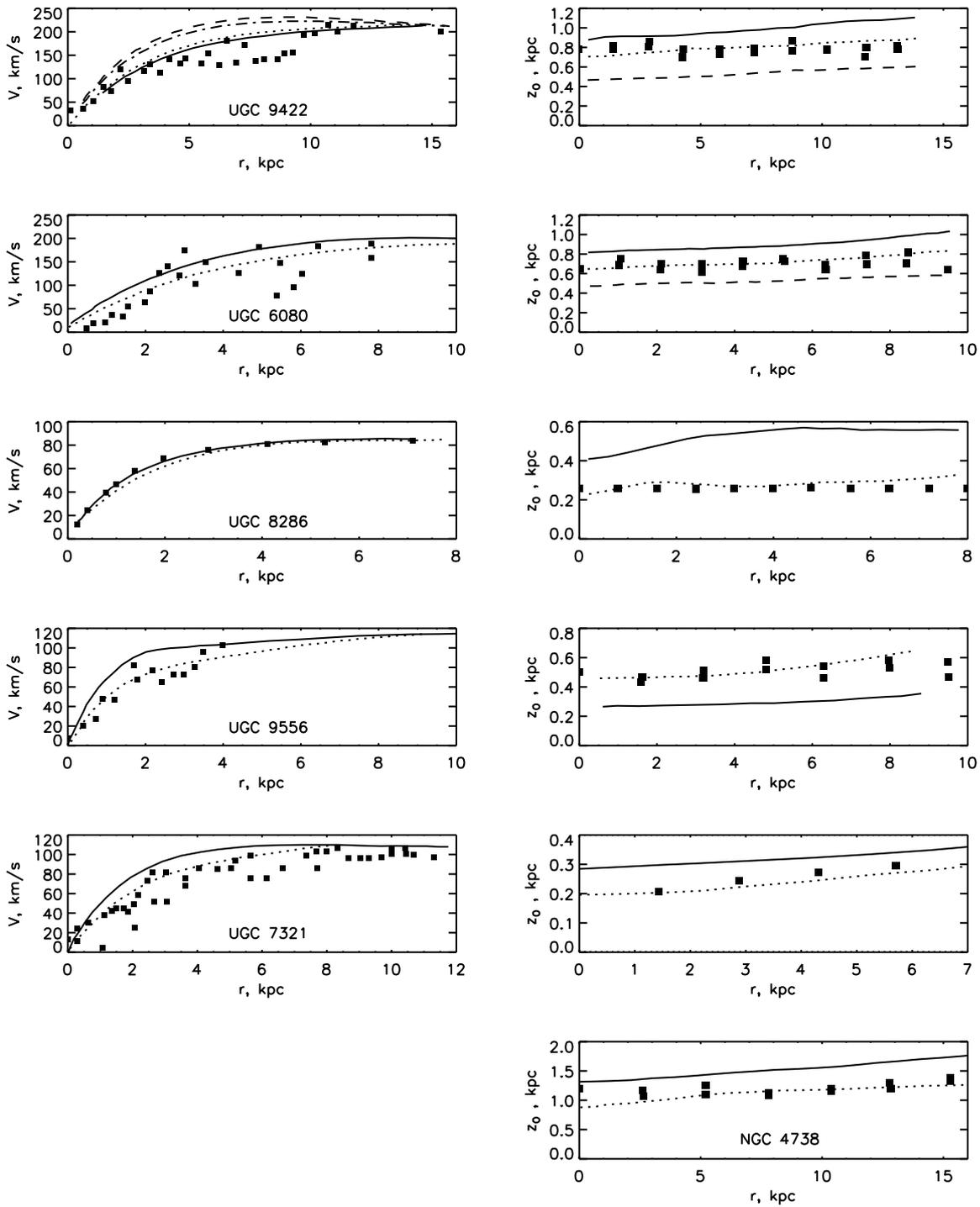}
\caption{
Comparison of our modelling with observed values for UGC~9422, UGC~6080,
UGC~8286, UGC~9556, UGC~7321 and NGC~4738 (from top to bottom).
{\it Left panels}:
radial distributions of $V_c$ (solid curves) and $V^{los}$,
i.e. observed rotation curves in edge-on galaxies (dotted curves).
{\it Right panels}:
radial distributions of the projected scale height. Curves correspond to
models with different $\mu$. See explanation of individual curves
in \S\ref{sec4.1}.
Filled squares designate observed values (RC or disc thickness) in all
panels.}
\label{fig10}
\end{figure*}

\clearpage
\begin{figure*}
\includegraphics[height=17cm,angle=270]{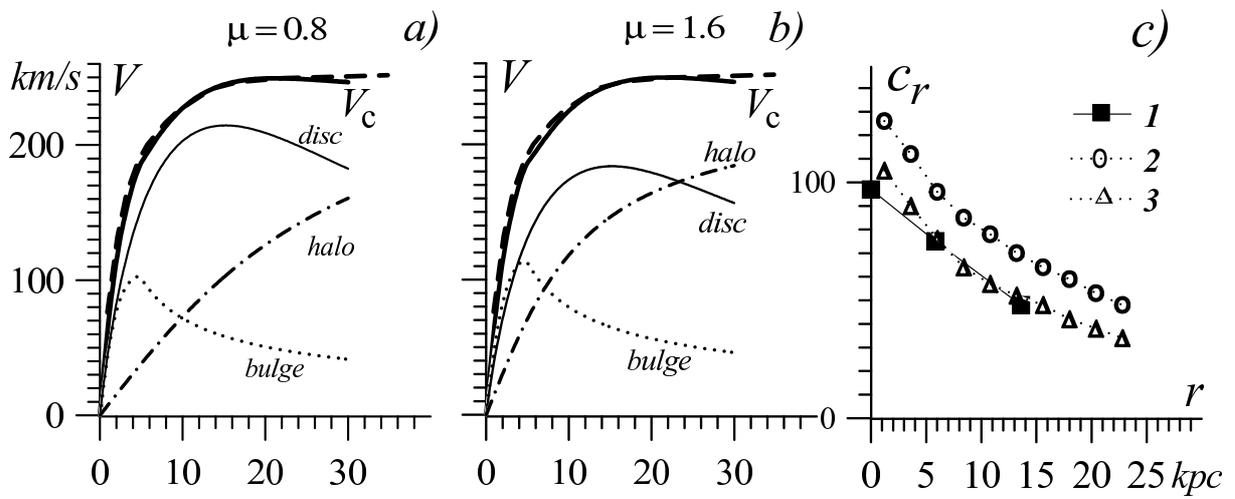}
\caption{Contributions of different galactic subsystems into the $V_c$
for two models of NGC~5170: $\mu=0.83$ (a)
and $\mu=1.6$ (b).
c) The radial velocity dispersions:
1 -- from observations by \cite{Bottema1987},
2 -- from the model with $\mu=0.83$ and
3 -- with $\mu=1.6$.
}
\label{fig11}
\end{figure*}


\clearpage
\begin{figure*}
\includegraphics[height=9cm]{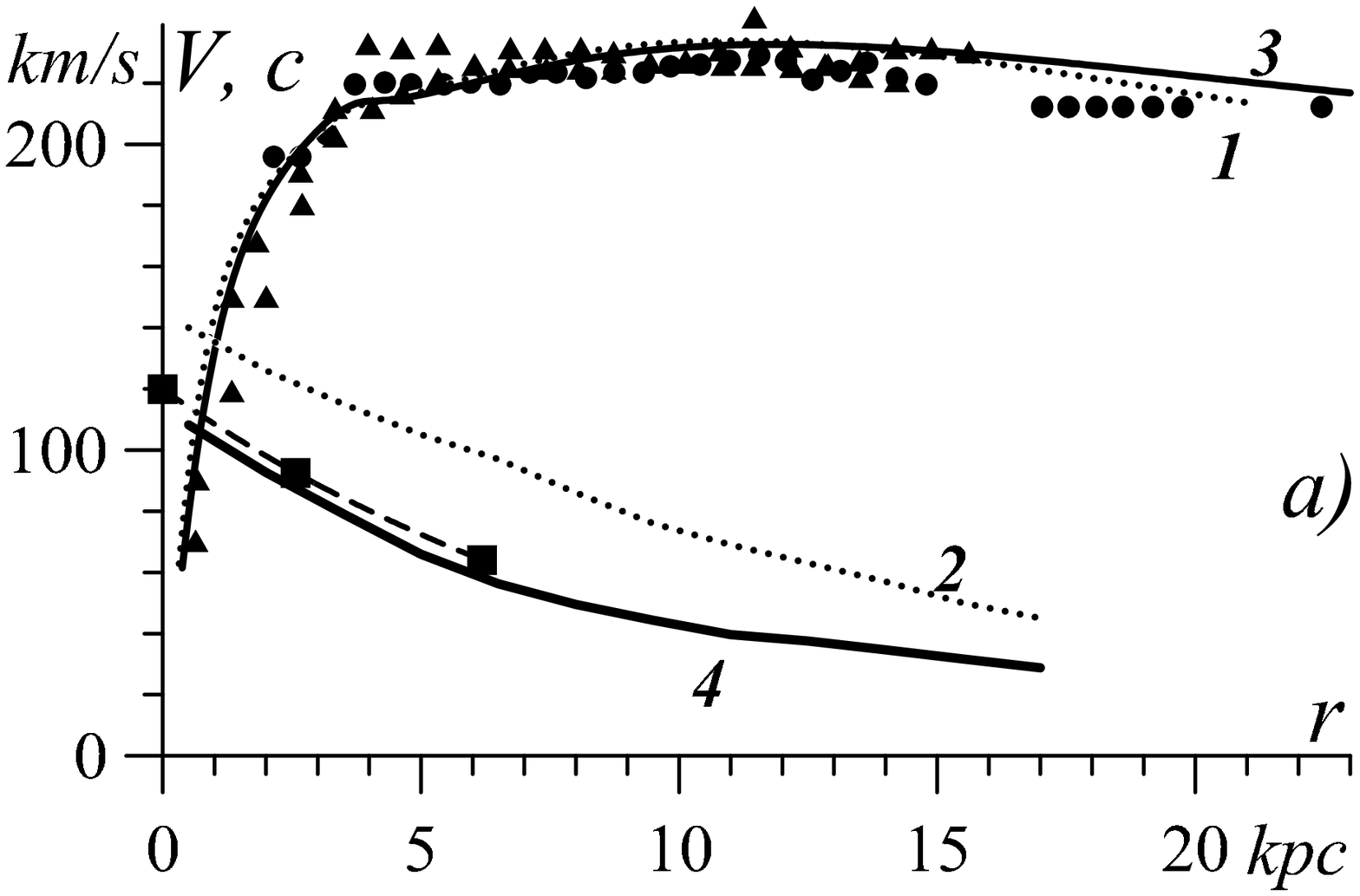}

\includegraphics[height=9cm]{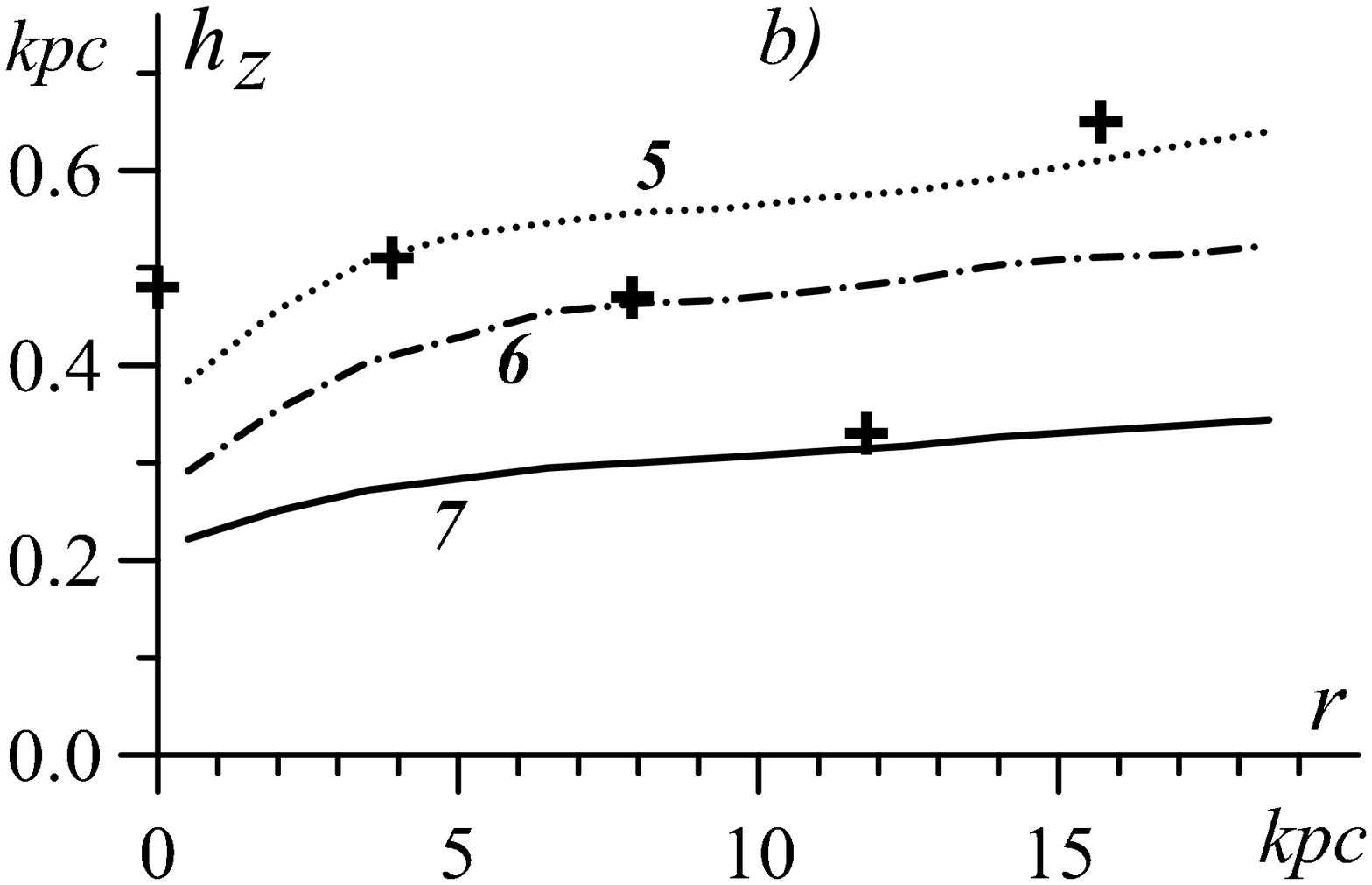}

\caption{NGC~891: a) Rotation curves from HI observations by
\cite{Bottema1991} (circles) and \cite{Sancisi1979} (triangles)
and the velocity dispersion from \cite{Bottema1991}. The model
distributions of $V_c$ ({\it 1}) and $c_r$ ({\it 2}) are shown for
the case of $M_h/M_d=0.46$, whereas the curves {\it 3} and {\it 4}
correspond to $V_c$ and $c_r$, respectively, for the case of
$M_h/M_d$ =1.5. b) Observed (crosses) and model (lines) projected
vertical scale height for the models with $M_h/M_d$ = 0.4 (the
curve 5), 0.7 (6) and 1.7 (7). } \label{fig12}
\end{figure*}

\clearpage
\begin{figure}
\includegraphics[width=7cm]{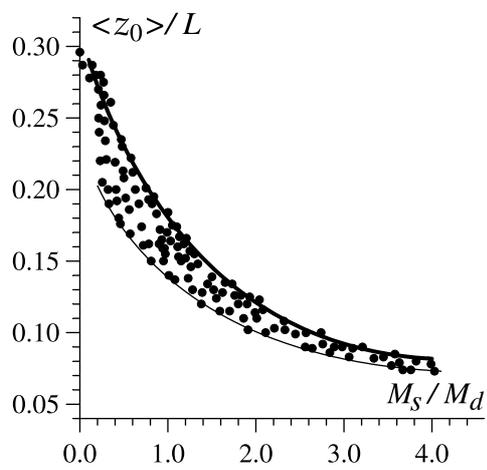}
\caption{The stellar disc thickness ($z_0/L$) and the
spherical-to-disc mass ratio $M_s/M_d$. The circles designate
models on the stability boundary with different parameters of
spherical subsystem and initial conditions in the modelling. The
upper curve corresponds to bulgeless models whereas the lower one
shows the models with noticeable bulges. } \label{fig14}
\end{figure}


\clearpage
\begin{figure}
\includegraphics[width=7cm]{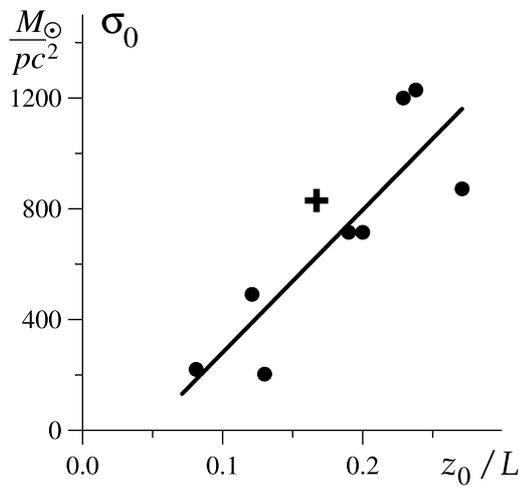}
\caption{The stellar disc thickness ($z_0/L$) and the disc central
surface density (in $M_{\odot}~pc^{-2}$) for the galaxies from our
sample (circles). The cross designates our Galaxy (\cite{AVHMW2003})
and the solid line is the least-square linear regression. }
\label{fig13}
\end{figure}


\clearpage
\begin{figure}
\includegraphics[width=7cm]{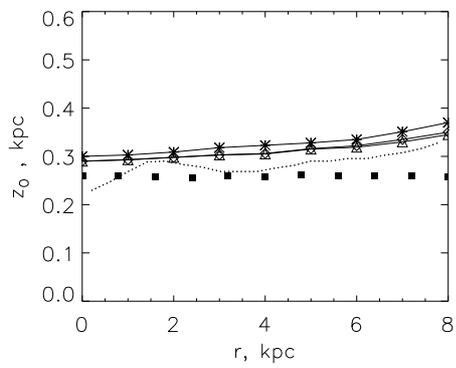}
\caption{
Results of "live" halo modelling for one of our galaxies, UGC~8286.
This figure is the same as the third upper and right panel in
Figure~\ref{fig10} with addition of three models which are designated by the
solid curves with asterisks, diamonds, and triangles, which correspond to
2$\cdot$10$^5$, 4$\cdot$10$^5$ and 6$\cdot$10$^5$ bodies in the "live" halo,
respectively.
}
\label{fig15}
\end{figure}


\end{document}